\documentclass{article}
\usepackage[margin=3cm]{geometry}
\usepackage[utf8]{inputenc}

\usepackage{amsmath,amssymb,amsfonts}
\usepackage{mathrsfs} 
\usepackage{mathtools}
\usepackage{todonotes}
\usepackage{graphicx,color}
\usepackage[nocompress]{cite}

\usepackage{authblk}

\numberwithin{equation}{section}

\usepackage{hyperref}
\hypersetup{colorlinks=true,allcolors=blue,urlcolor=cyan}

\title{Teleparallel Jackiw-Teitelboim gravity}

\author[1,2]{Christian G B\"ohmer\footnote{Email: c.boehmer@ucl.ac.uk}}
\author[3,4]{Rafael Ferraro\footnote{Email: ferraro@iafe.uba.ar}}
\author[5]{Franco Fiorini\footnote{Email: francof@cab.cnea.gov.ar}}

\affil[1]{Department of Mathematics, University College London, \authorcr 
Gower Street, London WC1E 6BT, UK\medskip}
\affil[2]{Astrophysics Research Centre, School of Mathematics, \authorcr 
Statistics and Computer Science, University of KwaZulu-Natal, \authorcr 
Private Bag X54001, Durban 4000, South Africa\medskip}
\affil[3]{Instituto de Astronomía y Física del Espacio (IAFE, CONICET-UBA), \authorcr
Buenos Aires, Argentina\medskip}
\affil[4]{Departamento de Física, Facultad de Ciencias Exactas y Naturales, \authorcr Universidad de Buenos Aires, Argentina\medskip}
\affil[5]{Grupo de Comunicaciones Ópticas, Departamento de Ingeniería en Telecomunicaciones, Consejo Nacional de Investigaciones Científicas y Técnicas (CONICET) and Instituto Balseiro (UNCUYO), Centro Atómico Bariloche,\authorcr
  Av.~Ezequiel Bustillo 9500, CP8400, S. C. de Bariloche, Río Negro, Argentina\medskip}
\date{\today} 

\begin{document}
\maketitle

\begin{abstract}
We introduce a new class of two dimensional gravity models using ideas motivated by the Teleparallel Equivalent of General Relativity. This leads to a rather natural formulation of a theory that has close links with Jackiw-Teitelboim gravity. After introducing the theory and discussing its vacuum solutions, we present the Hamiltonian analysis. This implies the presence of a single dynamical degree of freedom, which is in sharp contrast to General Relativity, where there are no degrees of freedom in two spacetime dimensions. Our approach can be extended to various other lower-dimensional gravity theories and thus could be of wider interest.
\end{abstract}

\clearpage

\section{Introduction}

Einstein's theory of General Relativity (GR) cannot be formulated directly in $1+1$ dimensions 
using either the Einstein-Hilbert action or the related
field equations. The Einstein-Hilbert action in $1+1$ dimensions is
topological, which means that the Ricci scalar can be written as a total derivative.
Consequently one cannot arrive at non-trivial field equations. On the other hand, it is also
well known that the $1+1$ dimensional Ricci tensor satisfies the simple geometrical
identity $R_{\mu\nu} = R g_{\mu\nu}/2$ which implies that the Einstein
tensor $G_{\mu\nu} := R_{\mu\nu} - R g_{\mu\nu}/2 \equiv 0$ vanishes
identically. This follows from the more general formula which allows one to
express the Riemann curvature tensor via the curvature scalar, namely
\begin{equation}  
    \label{Riem}
    R_{\mu\nu\lambda\rho}=\frac{1}{2}R\,(g_{\mu\lambda}\,g_{\nu\rho}-g_{\mu\rho}%
    \,g_{\nu\lambda}) \,.
\end{equation}
This identity only holds in two dimensions.

Moreover, all two dimensional manifolds $\mathscr{M}$ are conformally flat. This means, 
since the metric tensor in two dimensions has 3 independent
components and one can make two coordinate transformations, that the metric only
contains a single true degree of freedom. In suitably chosen local coordinates $%
(t,x)$ the line element or the metric can always be brought into the
following form
\begin{equation}  
    \label{met2din}
    ds^2 = \exp[2\Phi(t,x)](dt^2-dx^2) \,,
\end{equation}
where $\Phi(t,x)$ is the \emph{dilaton} field.

All of this means that one has to introduce additional fields or an
additional structure in order to formulate non-trivial gravity models in two spacetime dimensions. A
variety of models and different approaches were discussed in~\cite{Grumiller:2006rc} and references therein.\footnote{We can track allusions to 2D gravity to as far as the nineteen sixties and seventies, see \cite{six1,sev1}.} In particular, we could mention the scheme based on Riemann-Cartan geometry, where both curvature and torsion are dynamically incorporated into the gravitational action, see \cite{tanos,rusos}. Another approach is to consider a
two-dimensional singular limit of a higher-dimensional theory, see e.g.~\cite%
{Mann:1992ar} which has found renewed interest in the context of
Gauss-Bonnet type theories~\cite{Glavan:2019inb}. However, this has not been
without criticism~\cite{Arrechea:2020gjw,Ai:2020peo,Gurses:2020rxb}. Yet another
framework has recently been suggested in~\cite{Boehmer:2023lpb} where the Ricci
scalar was decomposed suitably to also allow for a two dimensional theory to
be formulated.

A particularly fruitful approach goes back to Jackiw~\cite{Roman1, Roman} and
Teitelboim~\cite{Claudio1, Claudio2, Claudio} (JT), who proposed a theory of
gravity in $1+1$ dimensions whose dynamical equation in vacuum is simply
\begin{equation}
    R-\Lambda =0 \,.  
    \label{JT1}
\end{equation}
When conformally flat metrics of the form~(\ref{met2din}) are considered, the
field equation~(\ref{JT1}) becomes
\begin{equation}
    \square\Phi=-\frac{\Lambda}{2}\exp[2\Phi]\,, \qquad \square = \eta^{\mu\nu}\nabla_\mu\nabla_\nu \,,
    \label{JT1LIU}
\end{equation}
which is Liouville's equation~\cite{Liouville}. Early works in the subject (see, e.g. \cite{BHT1,BHT}) showed that the theory admits black hole solutions with a consistently defined notion of entropy, very much alike its 4D-counterpart, and that it represents a rich toy model for quantum gravity \cite{He}. Nowadays, the main interest in such a theory
is linked to string theory. For a modern perspective, consult Ref. \cite{JTreview}.

While the theory is neat, an undesirable feature of the JT model is that its dynamics cannot
be derived from a variational principle whose (covariant) action depends
exclusively on the metric. As a matter of fact, Jackiw considered the
covariant action
\begin{equation}
    S\propto\int d^{2}x\,\sqrt{-g}\,N(R-\Lambda) \,,  
    \label{Jdil}
\end{equation}
which yields equation~(\ref{JT1}) when varying the action with
respect to the scalar field $N$. This field plays the role of a Lagrange
multiplier in the theory. 

However, when varying with respect to the metric, action~(\ref{Jdil}) produces the additional equation
\begin{equation}
    2 \nabla_{\mu}\nabla_{\nu} N + \Lambda g_{\mu\nu} N = 0 \,.  
    \label{tenscalar}
\end{equation}
In this way, Eqs.~(\ref{JT1}) or~(\ref{JT1LIU}) alone determine the metric's only free function
$\Phi$, while~(\ref{tenscalar}) subsequently determines $N$ with no restriction 
on $g_{\mu\nu}$. 

On the other hand, JT theory admits a Hamiltonian formulation similar to
that of higher dimensional theories, with the main difference that the
Lagrange multipliers that accompany the super-Hamiltonian and the
super-momentum -- the \emph{lapse function} $\eta^{\bot}$ and \emph{shift
vector} $\eta^{1}$, respectively, defined below -- are not variables to be varied. 
That is, the super-Hamiltonian and the super-momentum are not constraints, they generate the temporal and spatial deformations of the field configuration~\cite{Claudio1, Claudio2, Claudio}. Teitelboim's action, which is constructed entirely from the metric but is manifestly non-covariant, reads
\begin{equation}
    S\propto\int d^{2}x\,\Big[\Big(\Phi_{,\,0}-\eta^{1}\Phi_{,\,1}-\eta^{1}_{,%
    \,1}\Big)^2-\eta^{\bot}\Phi_{,\,1}^2-2\,\eta^{\bot}_{,\,1}\Phi_{,\,1}+%
    \Lambda\,\eta^{\bot}\exp [2\Phi]/2\Big] \,.
    \label{accchil}
\end{equation}
Here, the coordinates $(x^0,x^1)$ are such that the metric is written as
\begin{equation}
    g_{\mu\nu}=\exp[2\Phi]
    \begin{pmatrix}
    (\eta^{1})^2+(\eta^{\bot})^2 & \eta^{1} \\
    \eta^{1} & -1 
    \end{pmatrix} \,,
\end{equation}
which reduces to~(\ref{met2din}) when the fixed external fields are chosen as
$\eta^{1}=0$ and $\eta^{\bot}=1$. The fact that $\eta^{1}$ and $\eta^{\bot}$
are not considered as dynamical variables is reflected in the presence of a
central charge in the algebra of the generators.

As already mentioned, the Einstein-Hilbert action in two dimensions becomes trivial,
however, when working in the teleparallel setting, one can construct non-trivial terms for an action. This was recently done in Refs.~\cite{Andronikos} and~\cite{Tartu}, where several points associated to the role played by the breaking of the Lorentz symmetry were discussed. Here, we further examine this subject by performing a complete Hamiltonian analysis, which leads to a complete characterization of the additional degree of freedom present in the theory. In particular, we see that JT gravity emerges as a Lorentz invariant sub-sector of the 2D torsional action in consideration.   

Our notation is as follows: Latin indices $a,b,\ldots$ denote tangent space
indices, Greek indices $\mu,\nu,\ldots$ denote spacetime indices. The
tangent space metric is $\eta_{ab}$, the spacetime metric is $g_{\mu\nu}$,
we work with signature $(+,-)$. For explicit tangent space indices we use $%
a,b,\ldots = \underline{0}, \underline{1}, \ldots$, and drop the underline
for spacetime indices.

\section{Torsional action and field equations}
\label{sec2}

\subsection{The action}

Our first goal will be to show that JT dynamics can be obtained from a variational principle formulated in a teleparallel framework. To do this, let us rewrite the dynamical equation in terms of the exterior derivative of the vielbein or coframe 1-form $\mathbf{E}^{a} = E^a_\mu dX^\mu$. In any dimension, one defines the torsion 2-form to be
\begin{equation}
    \mathbf{T}^{a} := d\mathbf{E}^{a} \qquad \Longrightarrow \qquad
    T^{a}{}_{\mu \nu} = \partial _{\mu }E_{\nu }^{a}-\partial _{\nu }E_{\mu
    }^{a}= -T^{a}{}_{\nu \mu} \,.
\end{equation}
The vielbein has as its dual basis the frame vector fields $\mathbf{e}_{a}
= e_a^\mu \partial_\mu$ in the tangent space. These satisfy the relations
\begin{equation}
    e_{a}^{\mu } \, E_{\nu}^{a} = \delta _{\nu }^{\mu }\,, \qquad 
    e_{b}^{\mu } \, E_{\mu}^{a} = \delta _{b}^{a}\,,
\end{equation}
and are related to the metric tensor through
\begin{equation}
    g_{\mu \nu } = \eta _{ab}~E_{\mu }^{a}E_{\nu }^{b}\,, \qquad 
    \Leftrightarrow \qquad \eta_{ab}=g_{\mu \nu }~e_{a}^{\mu }e_{b}^{\nu } \,.
    \label{metric}
\end{equation}
In fact, the final relationship describes the orthonormality of the tangent space basis $\{\mathbf{e}_{a}\}$. This framework is said to be \textit{teleparallel} because $\mathbf{T}^{a}$ is the torsion of a zero (i.e., curvatureless) spin connection, also called a Weitzenböck connection. If the spin connection is zero, then the parallel transport of vectors become path-independent. Thus the manifold $\mathscr{M}$ is endowed with an absolute notion of parallelism. While Weitzenböck geometries are deprived of curvature, the torsion, instead, becomes the geometrical quantity that describes gravity. Beyond the interpretation of $d\mathbf{E}^{a}$ as a torsion field, our interest is focused on rewriting the JT equation (\ref{JT1}) in terms of second derivatives of the \textit{diad} or \textit{zweibein} $\{\mathbf{E}^{\underline{0}}, \mathbf{E}^{\underline{1}}\}$, note that in four dimensions one generally speaks of a \textit{tetrad} or \textit{vierbein}. 

For this purpose, we need to state the relationship between the curvature scalar $R$ of the Levi-Civita connection, which depends on second derivatives of the metric, and the first derivatives of torsion $T^{a}{}_{\mu \nu }$. As is well known, in any dimension this relation is (see, e.g. \cite{alpe})
\begin{equation}
    R=-T+ 2~E^{-1}\partial_{\rho}
    (E~T_{\nu }{}^{\nu \rho }) \,,  
    \label{R1}
\end{equation}
where $T=S_{\rho }{}^{\mu \nu }T^\rho{}_{\mu \nu }$, $E=\det E_{\mu }^{a} = |\det (g_{\mu \nu })|^{1/2}$ is the determinant of the co-frame components, and the torsion tensor in spacetime components is given by $T^{\rho }{}_{\mu\nu }=e_{a}^{\rho }\,T^{a}{}_{\mu \nu }$. The tensor $S^{\rho}{}_{\mu \nu }$ is defined by
\begin{equation}\label{elsen4}
    S^{\rho }{}_{\mu \nu }=\frac{1}{4}(T^{\rho }{}_{\mu \nu }-T_{\mu \nu
    }{}^{\rho }+T_{\nu \mu }{}^{\rho })+\frac{1}{2}(\delta _{\mu }^{\rho
    } T_{\sigma \nu }{}^{\sigma }-\delta _{\nu }^{\rho } T_{\sigma \mu
    }{}^{\sigma })=-S^{\rho }{}_{\nu \mu } \,,
\end{equation}
and is often referred to as the \emph{superpotential}. Since $S^{\rho}{}_{\mu \nu}$ is antisymmetric in the last pair of indices, in two dimensions the only independent components are $S^{\rho}{}_{01}$
\begin{align}
    S{}_{\rho 01} &= \frac{1}{4}(T_{\rho 01}-T_{01\rho }{}+T_{10\rho }{})+\frac{%
    1 }{2}(g_{0\rho }~g^{\lambda \sigma }T_{\sigma 1\lambda }-g_{1\rho
    }g^{\lambda \sigma }T_{\sigma 0\lambda }{})  \nonumber \\
    &= \frac{1}{4}(T_{\rho 01}-T_{01\rho }+T_{10\rho })+\frac{1}{2}T_{\sigma
    10} (g_{0\rho }~g^{0\sigma }{}+g_{1\rho }g^{1\sigma })  \nonumber \\
    &= \frac{1}{4}(T_{\rho 01}-T_{01\rho }+T_{10\rho })+\frac{1}{2}T_{\sigma
    10} \delta _{\rho }^{\sigma }=\frac{1}{4}(-T_{\rho 01}-T_{01\rho }+T_{10\rho
    }) \,.
\end{align}
Setting $\rho=0$ gives $-T_{001}-T_{010}+T_{100}=T_{010}-T_{010} = 0$, likewise for $\rho=1$ we find
$-T_{101}-T_{011}+T_{101}=-T_{101}+T_{101}=0$. Therefore, the superpotential vanishes identically 
when working in two dimensions. Thus, in two dimensions Eq.~(\ref{R1}) takes the form
\begin{equation}
    R=2 E^{-1}\partial_{\rho }(E~T_{\nu }{}^{\nu\rho})\,.  
    \label{R2}
\end{equation}
As we mentioned previously, this takes the form of a total derivative when multiplied by $E$ as it does for the Einstein-Hilbert action, thus leading to trivial field equations.

Let us now go back to the JT equation~(\ref{JT1}), which will resemble an electromagnetic equation if $R$ is replaced by the divergence of the vector $T_{\nu }{}^{\nu\rho}$, so that
\begin{equation}
    2 E^{-1} \partial_{\rho }(E T_{\nu }{}^{\nu \rho})-\Lambda =0 \,.  
    \label{JT2}
\end{equation}
It should be noted that this field equation is linear in the first derivatives of the torsion tensor. When field equations contain linear first derivatives, it is natural to seek an action which is quadratic in the relevant variables, generally called the field strengths. This is the case of electromagnetism, other Yang-Mills theories or elasticity theory.  This suggests to consider an action of the form\footnote{As an historical remark, we mention that (\ref{action}) is the 2D version of the \emph{Einstein's unified field theory}, an early tentative to unify gravitation and electromagnetism under the same geometrical setting. See \cite{Delp} for a compendium of works in the field.}
\begin{equation}
    S[\mathbf{E}^{a}]= k \int E 
    (\eta _{ab} T^{a}{}_{\rho \nu }T^{b \rho \nu} - \Lambda ) \, dx^{0} dx^{1} \,= k\int E 
    (\,\mathbb{T} - \Lambda ) \, dx^{0} dx^{1}\,,  
    \label{action}
\end{equation}
where we have defined $\mathbb{T}=\eta _{ab} T^{a}{}_{\rho \nu }T^{b \rho \nu}=T_{b \rho \nu }T^{b \rho \nu}$, and $k$ is an arbitrary constant which one can think of as the two dimensional coupling constant. This action will be our starting point for what follows, it will lead to non-trivial field equations and a gravitational toy model containing a true dynamical degree of freedom. This is a particularity of the torsion tensor in two dimensions. It contains two free functions and one can construct a unique scalar out of this quantity. Compared with GR and its standard formulation, the Riemann curvature tensor only contains one free function (determined by the metric) and we can again construct a unique scalar, the Ricci scalar. It just so happens that this scalar can be written as a total derivative term, hence making it unsuitable for an action principle.

\subsection{Field equations -- variations with respect to the diad}

To vary this action with respect to the diad, we recall the identity
\begin{equation}
    \delta E= E e_{a}^{\nu }~\delta E_{\nu }^{a} \,.
    \label{ddet}
\end{equation}
Next, we consider the quadratic torsion term
\begin{align}
   \delta(\mathbb{T}) =\delta (\eta _{ab}~T^{a}{}_{\rho \nu }T^{b\rho \nu }) &= 
    \delta (\eta_{ab} T^{a}{}_{\rho\nu} T^{b}{}_{\mu \lambda } g^{\rho \mu }
    g^{\nu\lambda})  \nonumber \\ &=
    2 \eta_{ab} \delta T^{a}{}_{\rho \nu} T^{b}{}_{\mu \lambda} g^{\rho
    \mu }g^{\nu \lambda } + 2\eta _{eb} T^{e}{}_{\rho \nu} 
    T^{b}{}_{\mu \lambda} g^{\rho \mu } \delta g^{\nu \lambda }  
    \nonumber \\ &=
    4 \eta _{ab} T^{b \rho \nu }\partial_{\rho }\delta E_{\nu }^{a}+
    4\eta_{eb} T^{e}{}_{\rho \nu } T^{b\rho}{}_{\lambda} \eta ^{cd}
    \delta e_{c}^{\nu } e_{d}^{\lambda} \,.  
    \label{variation}
\end{align}
We will now rewrite the second term
\begin{align}
    T^{e}{}_{\rho \nu } \eta ^{cd} \delta e_{c}^{\nu } e_{d}^{\lambda}
    &=
    T^{e}{}_{\rho \mu } \delta _{\nu }^{\mu } \eta ^{cd} 
    \delta e_{c}^{\nu} e_{d}^{\lambda } = 
    T^{e}{}_{\rho \mu } e_{a}^{\mu } E_{\nu }^{a} \eta^{cd} \delta e_{c}^{\nu } e_{d}^{\lambda}
    \nonumber \\ &= 
    -T^{e}{}_{\rho\mu} e_{a}^{\mu} \eta^{cd} e_{c}^{\nu} e_{d}^{\lambda} 
    \delta E_{\nu }^{a} = -T^{e}{}_{\rho\mu} e_{a}^{\mu } g^{\nu\lambda} \delta E_{\nu }^{a} \,.
\end{align}
Therefore, the complete variation of the quadratic term is 
\begin{equation}
     \delta(\mathbb{T}) = 
    4 \eta_{ab} T^{b \rho \nu }\partial_{\rho } \delta E_{\nu }^{a}-
    4 \eta_{eb} T^{e}{}_{\rho \mu } T^{b\rho \nu } e_{a}^{\mu } \delta E_{\nu }^{a} \,.
    \label{variation2}
\end{equation}
Combining this with~(\ref{ddet}) we arrive at the result
\begin{equation}
    \delta S \propto \int \Bigl[
    -4 \partial_{\rho }(E \eta _{ac} T^{c\rho \nu}) - 
    4E\eta _{eb}e_{a}^{\mu }T^{e}{}_{\rho \mu }T^{b\rho \nu} + 
    E e_{a}^{\nu } (\mathbb{T}-\Lambda )\Big] \delta E_{\nu }^{a} \,,
\end{equation}
where we left out a boundary term. The vacuum field equations of the theory are thus
\begin{equation}
    -4 \partial_{\rho }(E\eta _{ac}T^{c\rho \nu })-
    4E\eta_{eb} e_{a}^{\mu}\, T^{e}{}_{\rho \mu }T^{b\rho \nu }+
    E e_{a}^{\nu } (\mathbb{T}-\Lambda ) = 0 \,.  
    \label{fieldequations}
\end{equation}%
To show that these equations indeed contain the JT model, let us contract with $E_{\nu}^{a}$. 
By doing so we compute the trace of the field equations, namely
\begin{equation}
    -4 E_{\nu}^{a} \partial_{\rho}(E\eta _{ac}T^{c\rho \nu })-
    2E(\mathbb{T}+\Lambda)=0 \,.
\end{equation}%
Since in two spacetime dimensions we have $\mathbb{T}=2\eta _{ac}\partial_{\rho}E_{\nu}^{a}T^{c\rho\nu}$, 
the trace turns out to be
\begin{align}
    -4\partial_{\rho }(E\eta_{ac}E_{\nu}^{a}T^{c\rho\nu})-2E\Lambda &=0\, , 
    \nonumber \\
    \Leftrightarrow\qquad
    2 E^{-1}\partial_{\rho }(E E_{\nu}^{a}T_{a}{}^{\rho\nu})+\Lambda &=0\, ,  
    \nonumber \\
    \Leftrightarrow\qquad
    -2 E^{-1}\partial_{\rho }(E T_{\nu}{}^{\nu\rho})+\Lambda &=0 \label{jteq}
    \,.
\end{align}
As we had hoped, this is indeed Eq.~(\ref{JT2}), i.e., the JT model, derived from 
a natural and well-defined variational principle.\footnote{The $\Lambda=0$ version of Eq. (\ref{jteq}) was previously derived in Ref. \cite{Vatu2}.} Moreover, this did not 
require the introduction of somewhat arbitrary Lagrange multipliers.
By starting from an action that is not built from the metric but rather from 
the diad, we can provide a new approach to this model.

The fact that we are working in two dimensions allows for further
simplifications in the dynamical equations~(\ref{fieldequations}). Let us
consider the second term for each value of the index $\nu$. 
For $\nu=0$ one finds
\begin{equation}
    -4 E \eta _{eb} e_{a}^{\mu} T^{e}{}_{\rho\mu} T^{b\rho 0} =
    -4 E e_{a}^{0} \eta _{eb} T^{e}{}_{1 0} T^{b 1 0 } =
    -2 E e_{a}^{0}\, \mathbb{T} \,,
\end{equation}
and similarly for $\nu=1$. Thus the field equations~(\ref{fieldequations}) simplify to
\begin{equation}
    4 \partial _{\rho }(E\eta _{ac}T^{c\rho \nu })+Ee_{a}^{\nu }(\mathbb{T}+\Lambda)=0 \,.
    \label{fieldequations1}
\end{equation}
The two equations $\nu=1$ contain second time derivatives and can therefore be
seen as dynamic. Instead, the two equations $\nu=0$ constrain the initial values 
of the dynamical variables and their velocities.

\subsection{Solving the field equations}
\label{solve}

We begin with the conformally flat form of the metric tensor~(\ref{met2din}), whose scalar curvature is
\begin{equation}
    R=-2 \exp(-2\Phi)\,\square\Phi \,.
\end{equation}
Since the metric relates to the diad through Eq.~(\ref{metric}), we can use the diad
\begin{equation}
    \mathbf{E}^{{\underline{0}^\prime}} = \exp(\Phi) \mathbf{d}t 
    \,, \qquad
    \mathbf{E}^{{\underline{1}}^\prime} = \exp(\Phi) \mathbf{d}x \,.  
    \label{diad}
\end{equation}%
As the metric is invariant under local Lorentz transformations, there exists an
entire family of admissible diads. Let us consider a local Lorentz transformation 
(simply a boost, in $1+1$ dimensions), then the diads 
\begin{equation}
    \mathbf{E}^{a}=\Lambda _{~a^\prime}^{a}~(t,x)~\mathbf{E}^{a^\prime}\,, \qquad
    \Lambda _{~a^\prime}^{a}=%
    \begin{pmatrix}
    \cosh \phi (t,x) & \sinh \phi (t,x) \\
    \sinh \phi (t,x) & \cosh \phi (t,x)%
    \end{pmatrix} \,,
    \label{diad1}
\end{equation}%
constitute the entire set of diads giving the same conformally flat metric tensor (\ref{met2din}). Henceforth we denote by $\phi$ to the \textit{booston} field. Note that the field 
equations~(\ref{fieldequations1}) govern both the dilaton and the
booston fields. This follows since the action~(\ref{action}) 
depends on the diad and not just on the metric.\footnote{%
The action is invariant under global Lorentz and conformal transformations.} 

In particular, the dilaton field is directly obtained from the trace of the 
dynamical equations~(\ref{JT2}). Recall that this trace equation is the JT equation
and that it does not involve the booston because the curvature scalar $R$ 
depends only on the metric. The JT equation, then, comes out from this formalism as the local Lorentz invariant sector of the theory. The divergence $E^{-1} \partial_\rho(E T^\rho)$ of the vector
part of the torsion tensor
\begin{equation}
    T^{\rho} := T^{\nu}{}_{\nu }{}^{\rho} = g^{\rho \lambda } e_{a}^{\nu } T^a{}_{\nu\lambda}\,,  
    \label{vector}
\end{equation}
is also not sensitive to the booston, since both $R$ and $E$ in Eq.~(\ref{R2}) are invariant 
under local Lorentz transformations of the diad.

Therefore, $T^{\rho}$ can only depend on the booston through a divergence-free term.  
In fact, when computed from the diad~(\ref{diad1}) it results in\footnote{The respective covector is $T_\lambda dx^\lambda=-d\Phi+\star\, d\phi$. Besides, in action (\ref{action}) we have $E\,\mathbb{T}\, dx^0\wedge dx^1=2 (d\Phi-\star\, d\phi)\wedge \,\star\, (d\Phi-\star\, d\phi)$. In two dimensions, the Hodge star operator is not sensitive to the conformal factor; thus the dilaton is not present in $\star\, d\phi$.}
\begin{equation}
    E T^{\rho} \partial_\rho =\left(-\frac{\partial\Phi}{\partial t} +
    \frac{\partial\phi}{\partial x}\right)\partial_t +
    \left(\frac{\partial\Phi}{\partial x}-\frac{\partial\phi}{\partial t}\right)
    \partial_x\,,
    \label{vect}
\end{equation}
where we recall that $E=\exp(2\Phi)$. Once the dilaton is found by solving the trace equations, 
we will replace it in the rest of the equations to determine the booston. Thus
the action~(\ref{action}) provides dynamics to both of our variables: $\Phi$ and $\phi$.
The JT equation~(\ref{JT2}) for any diad is
\begin{equation}
    2\exp(-2\Phi) \square \Phi +\Lambda =0 \,.
    \label{dilaton}
\end{equation}
It turns out to be convenient to solve this equation using null coordinates. These
are given by
\begin{equation}
    u=\frac{t+x}{2}\,, \quad v=\frac{t-x}{2}\,, \qquad \Rightarrow \quad
    dt^{2}-dx^{2}=4 \,du\,dv \,.\label{nullcoord}
\end{equation}%
In these coordinates Eq.~(\ref{dilaton}) becomes
\begin{equation}
    \frac{\partial ^{2}\Phi }{\partial u\partial v}=-\frac{\Lambda }{2}\exp(2\Phi)~.
\end{equation}
Its general solution was given by Liouville in his seminal paper~\cite{Liouville}. It can be written in terms of two arbitrary \emph{chiral} functions, $U(u)$ and $V(v)$, such that
\begin{equation}
    \exp(2\Phi (u,v)) =\frac{U^{\prime}(u) V^{\prime }(v)}
    {\left[ 1+\frac{\Lambda }{2}~U(u)~V(v)\right]^{2}} \,.
    \label{dilatonuv}
\end{equation}
The freedom left in this solution does not reflect the existence of dilatonic waves, 
since such freedom can be completely absorbed in a change of chart, without altering the metric structure given in Eq.~(\ref{met2din}).
In fact, by writing the obtained metric in coordinates $(U,V)$, one obtains%
\begin{equation}
    ds^{2} =4 \exp(2\Phi)du\,dv = \frac{dU\,dV}{\left( 1+\frac{\Lambda }{2} (U-U_0)(V-V_0)\right)^{2}} \,.
    \label{metricUV}
\end{equation}
Here $U_0$ and $V_0$ are two arbitrary constants which can be eliminated by changing the origin of
coordinates.

One has to be somewhat careful about the interpretation of $\Phi$ as a true scalar field invariant under coordinate transformations. If we simply view $\Phi$ as the global pre-factor of the metric tensor, then clearly it cannot be seen as scalar as the metric picks up various contributions when making a coordinate transformation. This is clear when considering the transformation $(u,v) \mapsto (U,V)$, for the conformal factor has lost the piece $U^{\prime }(u)~V^{\prime }(v)$, compare Eqs.~(\ref{dilatonuv}) and~(\ref{metricUV}).\footnote{\label{foot}According to Eq.~(\ref{met2din}), $\Phi$ is a scalar under local Lorentz transformations of the chart $(t,x)$. Instead, $\Phi\rightarrow\Phi+1/2\log|u'(U)v'(V)|$ under transformations $u\rightarrow u(U)$, $v\rightarrow v(V)$. Thus, $\Phi$ does not change only if $u(U)=e^\zeta U$, $v(V)=e^{-\zeta} V$. } The conformal factor in the new chart $(U,V)$ implies that one should identify the dilaton as
\begin{equation}
    \Phi(U,V) = -\log \left[1+\frac{\Lambda }{2}~U V\right] \,.
    \label{dilatonUV}
\end{equation}
The coordinates $(U,V)$ can be transformed to Cartesian coordinates $(\tau ,\chi)$ taking a null-like form
\begin{equation}
    ds^{2}=\frac{d\tau ^{2}-d\chi ^{2}}{\left( 1+\frac{\Lambda }{8} 
    (\tau^{2}-\chi^{2})\right)^{2}}\,,  
    \label{metric-tau-chi}
\end{equation}
where $U=(\tau +\chi)/2$, $V=(\tau -\chi)/2$. Metric~(\ref{metric-tau-chi}) can also be 
associated with the diad
\begin{equation}
    \mathbf{E}^{{\underline{ 0}^\prime}} = 
    \frac{\mathbf{d}\tau }{1+\frac{\Lambda }{8} (\tau^{2}-\chi ^{2})}\,, \qquad
    \mathbf{E}^{{\underline{ 1}^\prime}} =
    \frac{\mathbf{d}\chi }{1+\frac{\Lambda }{8}~(\tau ^{2}-\chi ^{2})}\,,  
    \label{diad2}
\end{equation}
or any other diad linked to the previous one through a local Lorentz
transformation $\Lambda_{{a}^\prime}^{a}$ depending on the booston $\phi(\tau,\chi)$. 
Once the diad $\mathbf{E}^{a}=\Lambda_{{a}^\prime}^{a}(\phi)\mathbf{E}^{{a}^\prime}$ 
is substituted into Eqs.~(\ref{fieldequations}), the trace of the equations will be identically zero, 
since this diad is already associated with a metric satisfying the JT equation. 

The remainder of the field equations will determine the booston. It is useful to write 
the diad~(\ref{diad2}) in the chart $(U,V)$ which becomes
\begin{equation}
    \mathbf{E}^{{\underline{0}^\prime}}=\frac{\mathbf{d}(U+V)}{1+\frac{\Lambda }{2}~UV}\,, \qquad
    \mathbf{E}^{{\underline{1}^\prime}}=\frac{\mathbf{d}(U-V)}{1+\frac{\Lambda }{2}~UV}\,.
    \label{diad3}
\end{equation}
By replacing $\mathbf{E}^{a}=\Lambda_{{a}^\prime}^{a}(\phi) \mathbf{E}^{{a}^\prime}$ into the field 
equations (\ref{fieldequations1}), one obtains that $\phi$ must satisfy the equations
\begin{equation}
    \frac{\partial^2\phi}{\partial U\partial V}=0\,,\qquad
    \left(\frac{\partial\phi}{\partial U}\right)^2-\frac{\partial^2\phi}{\partial U^2} = 0 =
    \left(\frac{\partial\phi}{\partial V}\right)^2+\frac{\partial^2\phi}{\partial V^2} \,.
\end{equation}
These has the general solution
\begin{equation}
    \phi =\log \left|\frac{v_{0}\,V-\beta}{u_{0}\,U-\alpha}\right|  \,,
    \label{booston}
\end{equation}
with $u_0$, $v_0$, $\alpha$ and $\beta$ being constants of integration.

If the integration constants $u_0$ and $v_0$ are chosen to be zero, then $\phi$ becomes a global boost as it is now coordinate independent. In particular we find $\phi=0$ whenever $\alpha=\beta$. No boost means that the solution is the diad~(\ref{diad2})--(\ref{diad3}). However, in the general solution~(\ref{booston}) the booston is zero only along the straight lines $u_{0} U-\alpha=\pm(v_{0}V-\beta)$. Thus the integration constants represent the freedom to choose two (related) straight lines where the booston is zero.\footnote{$\alpha,\beta$ and $u_0,v_0$ cannot be absorbed into the coordinates $U,V$ because the scale and the coordinate origin were already fixed at the level of the dilaton.} Along these lines the field $\mathbf{E}^{a}=\Lambda_{{a}^\prime}^{a}(\phi) \mathbf{E}^{{a}^\prime}$ coincides with the diad $\mathbf{E}^{{a}^\prime}$ of Eq.~(\ref{diad3}). Therefore, the choice of the integration constants characterising the booston fixes a boundary condition for $\mathbf{E}^{a}$. 

For instance, the choice $\alpha=0=\beta$, $u_0=v_0 \neq 0$, implies that $\mathbf{E}^{a}$ will be equal to~(\ref{diad3}) along the straight lines $U=\pm V$, this means on the axes associated with the coordinates $\tau$, $\chi$. According to~(\ref{booston}) the function $\phi$ takes in this case the simple form
\begin{equation}
    \phi =\log \left|\frac{V}{U}\right|\,.
    \label{booston2}
\end{equation}
Using the two well-known identities $\cosh\log z=(z+z^{-1})/2$ and $\sinh\log z=(z-z^{-1})/2$, the Lorentz transformation~(\ref{diad1}) takes the form
\begin{equation}
    \Lambda_{{a}^\prime}^{a} = 
    \frac{1}{2}\begin{pmatrix}
        z + z^{-1} & z - z^{-1} \\
        z - z^{-1} & z + z^{-1}
    \end{pmatrix}\,,
    \qquad z = \frac{V}{U} \,.
\end{equation}
When this is put into the appropriate boosted diad, we find the following 
\begin{equation}
    \mathbf{E}^{{\underline{0}}} = 
    \frac{\frac{V}{U}~\mathbf{d}U+\frac{U}{V} \mathbf{d}V}{1+\frac{\Lambda }{2} UV} \,, \qquad 
    \mathbf{E}^{{\underline{1}} }=\frac{\frac{V}{U} \mathbf{d}U-\frac{U}{V} \mathbf{d}V}{1+\frac{\Lambda }{2} UV} \,.
    \label{vacuum}
\end{equation}
In this case, the booston will cover the entire range $(-\infty,\infty)$ in each quadrant of the
$(U,V)$ plane. In the coordinates $\tau,\chi$ the diad~(\ref{vacuum}) becomes
\begin{equation}
    \mathbf{E}^{{\underline{0}}} =
    \frac{(\tau^{2}+\chi^{2}) \mathbf{d}\tau -
    2\tau\chi \mathbf{d}\chi}{(1+\frac{\Lambda }{8} 
    (\tau ^{2}-\chi ^{2}))(\tau^2-\chi^2)} \,, \qquad
    \mathbf{E}^{{\underline{1}} } = 
    \frac{(\tau^{2}+\chi^{2})\mathbf{d}\chi -
    2 \tau\chi\mathbf{d}\tau}{(1+\frac{\Lambda }{8}
    (\tau ^{2}-\chi ^{2}))(\tau^2-\chi^2)} \,.
    \label{vacuum1}
\end{equation}
Leaving aside the conformal factor, the basis is $\pm (\mathbf{d}\tau,
\mathbf{d}\chi)$ on the Cartesian axes $\tau=0$ and $\chi=0$, while being
singular on the light cone $U=0$ or $V=0$ which corresponds to $(\tau+\chi)(\tau-\chi)=0$. 
The meaning of this solution can be more clearly understood by computing the torsion
vector (\ref{vector}). Notice that the diad in the action~(\ref{action}) can be seen as the
potential of the field $T^a{}_{\nu\lambda}$. The vector $T^{\rho}$ in two
dimensions has two independent components which matches the number of independent
components of the full torsion tensor $T^a{}_{\nu\lambda}$. This property is unique
to two dimensions, one can see this as follows. In $n$ dimensions the torsion tensor has
$n^2(n-1)/2$ independent components, for this to be equal to $n$ one has $n(n-1)/2=1$ or 
$n^2-n-2=0$, which has the unique positive solution $n=2$.

Therefore the torsion vector is sufficient to capture the nature of the entire
field. It should also be noted that $T^{\rho}$ is the dynamical field in the 
JT equation~(\ref{JT2}), which is perhaps not surprising now. The solution (\ref{diad2})--(\ref{diad3}),
which satisfied $\phi=0$, gives the torsion vector field
\begin{equation}
    T^{\rho} \partial_{\rho} = \frac{\Lambda}{4} \left(1+\frac{\Lambda}{2}UV\right) 
    \left(U\frac{\partial}{\partial U}+V\frac{\partial}{\partial V}\right) \,,
\end{equation}
where we recall that $T^{\rho}=0$ for Minkowski space using the trivial frame. 

On the other hand, the solution~(\ref{vacuum})--(\ref{vacuum1}) implies
\begin{equation}
    T^{\rho}\partial_{\rho} = -\frac{1}{2\,UV}\left(1+\frac{\Lambda }{2}UV\right) 
    \left(U\frac{\partial}{\partial U}+V\frac{\partial}{\partial V}\right) \,.
\end{equation}
The field lines of $T^{\rho}$ are radial with respect to the origin of coordinates, and 
$T^{\rho}$ is singular on the light cone $U=0$ or $V=0$. If $\alpha,\beta$ were different 
from zero, and $u_0=v_0=1$ to keep the slope of the lines fixed where $\phi=0$, 
all these field lines  would be centered at $(U,V)=(\alpha,\beta)$. This matches our
previous discussion of the meaning of the integration constants.

\section{Hamiltonian formalism}

In the following we will perform the Hamiltonian analysis of the action~(\ref{action}) and demonstrate that this theory contains a single dynamical degree of freedom. To begin, we compute the canonical momenta
\begin{equation}
    \pi_{a}^{\mu } := \frac{\partial L}{\partial (\partial _{0}E_{\mu }^{a})}%
    =\frac{\partial L}{\partial T^{a}{}_{0\mu }} \,,  
    \label{momentum}
\end{equation}
where $L$ is the Lagrangian density, as implied by action~(\ref{action}). Making the time derivatives explicit yields
\begin{align}
    L/k &= E (\eta_{ab} T^{a}{}_{\rho \nu  }T^{b\rho \nu }-\Lambda) =
    2E \eta_{ab}T^{a}{}_{01}T^{b01}-E \Lambda  
    \nonumber \\ &= 
    2 E \eta_{ab} T^{a}{}_{01} g^{0\rho } g^{1\lambda } T^{b}{}_{\rho\lambda }-E \Lambda 
    \nonumber \\ &= 
    2 E \eta_{ab} (g^{00}g^{11}-g^{01}g^{10}) T^{a}{}_{01}T^{b}{}_{01}-E \Lambda
    \nonumber \\ &= 
    -2E^{-1} \eta_{ab}~T^{a}{}_{01}T^{b}{}_{01}-E \Lambda \,.
    \label{Lag}
\end{align}
Since this Lagrangian does not contain the generalised velocity $\partial_{0}E_{0}^{a}$ (%
$E_{0}^{a}$ is deprived of dynamics), two \textit{primary} constraints will
appear. The momenta~(\ref{momentum}) conjugated to $E_{0}^{a}$ are zero
\begin{equation}
    G_{a}^{(1)} := \pi _{a}^{0} = 0 \,.  
    \label{primary}
\end{equation}%
Due to the fact that the Lagrangian contains time derivatives of $E_{1}^{a}$, we find that its conjugated momenta are
\begin{equation}
    \pi_{a}^{1} = -\frac{4k}{E} \eta_{ab} T^{b}{}_{01} = -\frac{4k}{E} T_{a01}\,.  
    \label{pia1}
\end{equation}
This relation shows, in passing, that parts of the torsion tensor are the conjugated momenta of this theory.
This is of course expected since torsion contains the first derivatives of the diads which are the dynamical
variables of the theory. It also shows that the torsion plays an important role in the 
Hamiltonian analysis of any teleparallel-formulated theory of gravity.

The next step is to state the primary Hamiltonian density, which is given by
\begin{align}
    H_{P} &= H_C+\lambda ^{b} \pi _{b}^{0} =
    \pi _{b}^{1}~\partial_{0}E_{1}^{b}-L+\lambda ^{b} \pi _{b}^{0} 
    \nonumber \\ &=
    \pi_{b}^{1} (T^{b}{}_{01}+\partial _{1}E_{0}^{b})-L+\lambda ^{b} \pi _{b}^{0} \,,
\label{pHam}
\end{align}%
where we used the definition of the torsion tensor in the final step. $H_C$ is the 
canonical Hamiltonian, and the Lagrange multipliers $\lambda ^{a}(t,x)$ account for 
the dynamically undetermined velocities $\partial _{0}E_{0}^{a}$. 
By substituting $T^{a}{}_{01}=-E/(4k) \eta^{ab} \pi _{b}^{1}$ into~(\ref{pHam}), 
the primary Hamiltonian density turns out to be
\begin{equation}
    H_{P}=-\frac{E}{8k}~\eta ^{cb} \pi _{c}^{1}\pi _{b}^{1}+\pi_{b}^{1} 
    \partial_{1}E_{0}^{b}+kE\Lambda +\lambda^{b}\pi_{b}^{0} \,.
\label{Hamp}
\end{equation}

In the Dirac-Bergmann formalism for constrained Hamiltonian systems, the
primary constraints are subjected to be consistent with the evolution. Sometimes
this condition is achieved through a suitable choice of the Lagrange
multipliers $\lambda^{a}$. If this is not the case, additional (secondary)
constraints should be imposed. The knowledge of the complete constraint
algebra allows the determination of the number of genuine degrees of freedom
and the gauge freedom of the system. 

In the Hamiltonian formalism this consistency condition takes the form
\begin{equation}
    \frac{d}{dt}G_{a}^{(1)}(t,x)=:\dot{G}_{a}^{(1)}(t,x)=
    \dot{\pi}_{a}^{0}(t,x)=\{\pi _{a}^{0}(t,x),\int dx^{\prime } H_{P}(t,x^{\prime })\}=0 \,,
\end{equation}%
where $\{,\}$ stands for the Poisson bracket defined by
\begin{equation}
    \{A(x),B(x^{\prime })\}=\int dy~\left( \frac{\delta A(x)}{\delta E_{\mu
    }^{b}(y)}\frac{\delta B(x^{\prime }))}{\delta \pi _{b}^{\mu }(y)}-\frac{%
    \delta A(x)}{\delta \pi _{b}^{\mu }(y)}\frac{\delta B(x^{\prime })}{\delta
    E_{\mu }^{b}(y)}\right) \,.
\end{equation}%
A dot above a quantity will denote its time derivative. Then
\begin{equation}
    \dot{G}_{a}^{(1)}(t,x) = -\int dy~\delta (x-y)~\frac{\delta }{\delta
    E_{0}^{a}(y)}\int dx^{\prime }~H_{P}(t,x^{\prime })=0 \,.
    \label{consistency1}
\end{equation}%
To be completely general, let us compute the variation of the Hamiltonian with
respect to each component $E_{\mu }^{a}$. According to Eq.~(\ref{Hamp}), $%
H_{P}$ depends on $E_{\mu }^{a}$ through its determinant $E$. We will use
Eq.~(\ref{ddet}) to compute $\delta E/\delta E_{\mu }^{a}$. Moreover $H_{P}$
depends on $E_{0}^{a}$ through the term $\pi _{b}^{1} \partial _{1}E_{0}^{b}$. 
Therefore we arrive at
\begin{align}
    \frac{\delta }{\delta E_{\mu }^{a}(y)}\int dx^{\prime }H_{P}(x^{\prime }) &= 
    e_{a}^{\mu }(y) E(y)~\frac{\delta }{\delta E(y)}\int dx^{\prime
    }H_{p}(x^{\prime })+\delta _{0}^{\mu }~\frac{\delta }{\delta E_{0}^{a}(y)}%
    \int dx^{\prime }~\pi _{b}^{1}(x^{\prime })~\partial _{1}E_{0}^{b}(x^{\prime
    })  
    \nonumber \\[1ex] &= 
    \int dx^{\prime }~e_{a}^{\mu } E~(-\frac{1}{8k}~\eta ^{dc} \pi_{d}^{1}
    \pi _{c}^{1}+k~\Lambda ) \delta (x^{\prime }-y)+\delta _{0}^{\mu
    }\int dx^{\prime } \pi _{a}^{1}(x^{\prime })~\partial _{x^{\prime }}\delta
    (x^{\prime }-y)  
    \nonumber \\[1ex] &=
    \left[ e_{a}^{\mu } E~(-\frac{1}{8k}~\eta ^{dc} \pi _{d}^{1}\pi_{c}^{1}+k~\Lambda )-
    \delta _{0}^{\mu }~\partial _{1}\pi _{a}^{1}\right]_{y} \,.  
    \label{varHp}
\end{align}
Thus the consistency condition~(\ref{consistency1}) for the evolution of the
primary constraints forces the introduction of secondary constraints because 
$\dot{G}_{a}^{(1)}$ does not vanish identically. Hence, we define $G_{a}^{(2)}$
to be
\begin{equation}
    G_{a}^{(2)} := e_{a}^{0}~E~(-\frac{1}{8k}~\eta ^{dc}\pi _{d}^{1}\pi_{c}^{1}+k\Lambda )-
    \partial _{1}\pi _{a}^{1} = 0 \,.
    \label{secondary}
\end{equation}
We see that the secondary constraints are indeed a part of the Lagrangian dynamics, Eqs.~(\ref{fieldequations1}) for $\nu=0$. In
fact, the field equations for $\nu =0$ are
\begin{equation}
    \partial_{1}\frac{\partial L}{\partial (\partial _{1}E_{0}^{a})}-
    \frac{\partial L}{\partial E_{0}^{a}} = 0 \,,
    \label{nueq0}
\end{equation}%
because $\partial_{0}E_{0}^{a}$ is absent in $L$. Since the derivatives of
the diad in $L$ appear in the antisymmetric combination $\partial
_{0}E_{1}^{a}-\partial _{1}E_{0}^{a}$, one verifies that
\begin{equation}
    \partial_{1}\frac{\partial L}{\partial (\partial _{1}E_{0}^{a})} = 
    -\partial_{1}\frac{\partial L}{\partial (\partial _{0}E_{1}^{a})} =
    -\partial _{1}\pi_{a}^{1} \,.
\end{equation}%
In addition, by using results from Section \ref{sec2} we obtain
\begin{equation}
    \frac{\partial L}{\partial E_{0}^{a}}=-k~E~e_{a}^{0}~(\eta
    _{bc}~T^{b}{}_{\rho \lambda }T^{c~\rho \lambda }+\Lambda
    )=-e_{a}^{0}~(L+2k~E~\Lambda ) \,,
    \label{L0}
\end{equation}%
which coincides with the second term in Eq.~(\ref{secondary}) once the
momenta~(\ref{pia1}) are substituted back into the Lagrangian~(\ref{Lag}). Thus the
solutions of Section~\ref{solve} satisfy the constraints~(\ref{secondary}).

In electromagnetism, for example, the secondary constraint is Gauss' law. In
teleparallel gravity the term $\partial L/\partial E_{0}^{a}$ additionally
appears in the constraint~(\ref{secondary}). This happens because
teleparallel Lagrangians, even if they resemble the electromagnetic
Lagrangian, depend not only on the derivatives of the vielbein but on the
vielbein itself. Nonetheless, we can recombine the two constraints $G_{a}^{(2)}$
to separate Gauss' law from the other contributions. Let us define $G_{\nu
}^{(2)} := E_{\nu }^{a}~G_{a}^{(2)}$, so that
\begin{equation}
    G_{\nu }^{(2)} = \delta _{\nu }^{0} E(-\frac{1}{8k} \eta^{dc}\pi _{d}^{1}
    \pi_{c}^{1}+k\Lambda )-E_{\nu }^{b}~\partial _{1}\pi _{b}^{1} \,.
    \label{super}
\end{equation}%
In the above, $G_{0}^{(2)}$ and $G_{1}^{(2)}$ are respectively the \textit{super-Hamiltonian}
and the \textit{super-momentum} constraints, using the familiar ADM language. In particular,
$G^{(2)}_0$ is equal to
\begin{equation}
    G^{(2)}_0=H_C-\partial_1(E^a_0\pi^1_a) \,.  
    \label{superH}
\end{equation}
To complete the analysis, we need to check further consistency conditions.

\subsection{Consistency of the secondary constraints}
\label{consis}

The Dirac-Bergmann algorithm is not complete until all the constraints are
consistent with the evolution. So, we have to impose the consistency of $%
G_{a}^{(2)}$, this means $\dot{G}_{a}^{(2)}$ must be zero at least on the constraint
surface. We say it is \textit{weakly} zero: $\dot{G}_{a}^{(2)}\approx 0$. 

Before computing $\dot{G}_{a}^{(2)}$, let us investigate what the Lagrangian
dynamics imply. Consider the divergence of the Euler-Lagrange equations
\begin{equation}
    \partial _{\nu }\partial _{\rho }\frac{\partial L}{\partial (\partial _{\rho}E_{\nu }^{a})}-
    \partial _{\nu }\frac{\partial L}{\partial E_{\nu }^{a}} = 0 \,,
\end{equation}%
where the first term is identically zero since the operator $\partial _{\nu}\partial _{\rho }$ 
is symmetric but the dependence of $L$ on $\partial_{\rho }E_{\nu }^{a}$ is antisymmetric. 
Then the solutions to the Euler-Lagrange equations satisfy the equations
\begin{equation}
    0=\partial _{\nu }\frac{\partial L}{\partial E_{\nu }^{a}}=\partial _{0}%
    \frac{\partial L}{\partial E_{0}^{a}}+\partial _{1}\frac{\partial L}{%
    \partial E_{1}^{a}}\,.
\end{equation}
As was mentioned below Eq.~(\ref{L0}), $\partial L/\partial E_{0}^{a}=-G_{a}^{(2)}-\partial _{1}\pi _{a}^{1}$. 
Moreover, on-shell we have 
\begin{equation}
    \frac{\partial L}{\partial E_{1}^{a}} = 
    \partial_{\rho} \frac{\partial L}{\partial (\partial_{\rho } E_{1}^{a})} =
    \partial_{0} \frac{\partial L}{\partial (\partial_{0}E_{1}^{a})} =
    \partial_{0} \pi_{a}^{1} \,. 
\end{equation}
Therefore one can arrive at the desired result
\begin{equation}
    0=-\partial _{0}(G_{a}^{(2)}+\partial _{1}\pi _{a}^{1})+
    \partial_{1}\partial_{0}\pi _{a}^{1} 
    \qquad \Rightarrow \qquad
    \partial_{0}G_{a}^{(2)}=0 \,.
\end{equation}
We can conclude that the Lagrangian dynamics implies the consistent evolution of the secondary
constraints, meaning that no further constraints will appear. 

Let us check that very same result using the Hamiltonian approach to compute 
the time evolution $\dot{G}_{a}^{(2)}$. We have
\begin{equation}
    \dot{G}_{a}^{(2)}=\{{G}_{a}^{(2)},\int dx^{\prime }H_{C}\}+\{{G}%
    _{a}^{(2)},\int dx^{\prime }~\lambda ^{c}\pi _{c}^{0}\} \,.
    \label{secondary2}
\end{equation}%
First we will consider the second Poisson bracket in Eq.~(\ref%
{secondary2}) and check whether the Lagrange multipliers $\lambda ^{c}$ will
be determined by the consistency condition $\dot{G}_{a}^{(2)}\approx 0$, which gives
\begin{equation}
    \{G_{a}^{(2)},\int dx^{\prime }~\lambda ^{c}\pi _{c}^{0}\}=\{e_{a}^{0}E~(-%
    \frac{1}{8k}~\eta ^{dc}\pi _{d}^{1}\pi _{c}^{1}+k\Lambda )~,\int dx^{\prime
    } \lambda ^{c}\pi _{c}^{0}\} \,.
\end{equation}%
In two dimensions we have the simple identity
\begin{equation}
    e_{b}^{\lambda }E = \epsilon_{bc} \epsilon^{\lambda\rho} E_{\rho }^{c} \,.\label{2Didentity}
\end{equation}%
This implies the simple result $\{e_{a}^{0}E,\pi _{d}^{0}\}=\epsilon _{bc}\{E_{1}^{c},\pi
_{d}^{0}\}=0$. Therefore, no Lagrange multipliers are present in the
consistency condition, which turns out to be
\begin{equation}
    0\approx \{{G}_{a}^{(2)},\int dx^{\prime }H_{C}\}=\{e_{a}^{0}\left(
    H_{C}-\pi _{b}^{1}~\partial _{1}E_{0}^{b}\right) -\partial _{1}\pi
    _{a}^{1}~,\int dx^{\prime }H_{C}\} \,,
    \label{evolveGa}
\end{equation}
where we have chosen for $G_{a}^{(2)}$ a convenient form to compute the
Poisson bracket, since 
\begin{equation}
    \{H_{C},\int dx^{\prime }H_{C}\}=0 \,, 
\end{equation}
and recall that $H_{C}$ does not depend on $\pi_{c}^{0}$. The proof that Eq.~(\ref{evolveGa}) 
is satisfied on the constraint surface is left to the Appendix~\ref{Ga}, as
the complete calculation is rather cumbersome.

Consequently, the Dirac-Bergmann algorithm has terminated without determining the Lagrange
multipliers $\lambda^c(t,x)$ taking part in the Hamiltonian $H_P$ in the terms $\lambda^c \pi^0_c$. 
This means that the evolution of the variables $E_{0}^{a}$ is left undetermined by the 
Hamiltonian $H_{P}$. Therefore these are pure gauge variables, similar to what happens to the 
component $A_0$ of the electromagnetic potential. This also means that the $E_{0}^{a}$ in 
the solution of Section~\ref{solve} has been gauge fixed.

\subsection{Symmetries of the action}

General relativity is covariant under coordinate transformation or diffeomorphisms. 
When considering infinitesimal coordinate transformations, the metric transforms
according to $\mathbf{g} \mapsto \mathbf{g}+ \mathcal{L}_{\mathbf{\xi }}\mathbf{g}$, 
where $\mathcal{L}_{\mathbf{\xi }}$ is the Lie derivative along the infinitesimal vector $\mathbf{\xi }$. This form is a common feature for all the theories displaying an
invariance under reparametrisations. 

In the Arnowitt-Deser-Misner (ADM) Hamiltonian formulation of GR, the
components $g_{0\mu}$ are not canonical variables, they play the role of
Lagrange multipliers, the lapse function and the shift vector. 
The canonical variables are the $d(d-1)/2$ components of the spatial block of
the metric where $d$ is the dimension of space-time. In standard GR one finds $6$
canonical variables. The space-time translations of $^{(d-1)}\mathbf{g}$ are 
generated by the super-Hamiltonian and super-momenta constraints. These 
constraints are 1st class, each commutes with all constraints. 

According to the Dirac-Bergmann formalism, 1st class constraints generate gauge
transformations, each one reveals the presence of a spurious degree of
freedom (d.o.f.) among the canonical variables. Since there are $d-1$
super-momenta constraints plus one super-Hamiltonian constraint, they
eliminate $d$ d.o.f. Therefore $^{(d-1)}g_{ij}$ contains just $d(d-1)/2-d=d(d-3)/2$
genuine d.o.f. Thus, GR has no (local) d.o.f. in $d=3$, since the dynamical
equations $R_{\mu\nu}=0$ leaves no room for a non-zero Riemann tensor. It is well-known that the Riemann tensor in $d=3$ is completely determined by the Ricci tensor.

The teleparallel formulation of Jackiw-Teitelboim theory also displays the
invariance under space-time translations. Let us first notice that the
Lagrangian~(\ref{Lag}) can be written as
\begin{equation}
    L = k E \left( 
    \frac{1}{2} f_{ce}^{\,\,\,\,a} f_{df}^{\,\,\,\,b}~M_{ab}{}^{cdef}-\Lambda
    \right) \,,
    \label{Lagf}
\end{equation}%
where $f_{ce}^{\,\,\,\,a}$ are the anholonomy coefficients,
\begin{equation}
    [\mathbf{e}_{a},\mathbf{e}_{b}] = f_{ab}^{\,\,\,\,c} \,\mathbf{e}_{c} \,.
\end{equation}
Moreover, the \textit{supermetric} $M_{ab}{}^{cdef}$ is given by
\begin{equation}
    M_{ab}{}^{cdef} = 2 \eta_{ab} \eta^{c[d} \eta ^{f]e}\,.
\end{equation}
The brackets $[]$ stand for skew-symmetrization over the indices. In fact, we have
\begin{equation}
    f_{ab}^{\,\,\,\,c} = E_{\lambda }^{c} 
    (e_{a}^{\rho }\partial_{\rho }e_{b}^{\lambda}-
    e_{b}^{\rho }\partial_{\rho }e_{a}^{\lambda }) = 
    -2e_{a}^{\rho}e_{b}^{\lambda }\partial_{[ \rho }E_{\lambda ]}^{c} = 
    -e_{a}^{\rho}e_{b}^{\lambda }~T^{c}{}_{\rho \lambda } \,,
\end{equation}
so that one can verify
\begin{equation}
    f_{ce}^{\,\,\,\,a} f_{df}^{\,\,\,\,b} M_{ab}{}^{cdef} = 
    e_{c}^{\rho }e_{e}^{\lambda} T^{a}{}_{\rho \lambda }
    e_{d}^{\mu}e_{f}^{\nu } T^{b}{}_{\mu\nu} 2\eta_{ab} \eta^{c[d}\eta^{f]e} =
    \eta _{ab} T^{a}{}_{\rho \lambda } T^{b}{}_{\mu\nu } 
    2g^{\rho [\mu }g^{\nu ]\lambda }=2\eta _{ab} T^{a}{}_{\rho
    \lambda } T^{b \rho \lambda}\,.
\end{equation}%
It is easy to prove that the Lagrangian is (pseudo) invariant under
space-time translations of the diad, namely
\begin{equation}
    \mathbf{E}^{a}\mapsto \mathbf{E}^{a}+\mathcal{L}_{\mathbf{\xi }}%
    \mathbf{E}^{a} \,.  
    \label{trans}
\end{equation}
Using the previous result from~\cite{Ferraro}, we have
\begin{equation}
    \delta_{\mathbf{\xi }} \mathbf{E}^{a} =
    \mathcal{L}_{\mathbf{\xi }}\mathbf{E}^{a} 
    \qquad\Rightarrow\qquad
    \delta_{\mathbf{\xi }} f_{bc}^{\,\,\,\,a}=\xi^{\nu }\partial _{\nu }f_{bc}^{\,\,\,\,a}\,,
    \qquad \delta_{\mathbf{\xi }} E=\partial_{\nu }(E\xi ^{\nu }) \,.
\end{equation}
We thus obtain that the change of the Lagrangian~(\ref{Lagf}) is
\begin{equation}
    \delta_{\mathbf{\xi }} L = 
    k \,\partial_{\nu }\left[ E\xi ^{\nu }\left( \frac{1}{2} f_{ce}^{\,\,\,\,a} f_{df}^{\,\,\,\,b} M_{ab}{}^{cdef}-\Lambda \right) \right] = 
    \partial_{\nu }\left[ \xi ^{\nu }L\right] \,.
\end{equation}
Then, the Lagrangian is \textit{pseudo}-invariant under space-time translations 
of the diad. By this we mean invariant up to a divergence. So, the field
equations have a \textit{local} invariance, since the functions $\xi^\nu(t,x)$ 
are arbitrary, and the translation~(\ref{trans}) maps a solution into
another solution for all infinitesimal vector fields $\mathbf{\xi}$.
Since any solution has curvature scalar $R$ equal to $\Lambda$, one
concludes that the space-time translations preserve the geometry, at first
order in $\mathbf{\xi}$. This is expected since the two-dimensional geometry is
completely characterized by $R$. Therefore, the translated solution corresponds
to the same geometry using a different diad, and we emphasise that the 
chart is not being changed however the diad is. At this point we wonder 
whether the theory can distinguish these two related solutions, or whether 
they differ in pure gauge variables only. This means we are interested in 
understanding whether or not those changes of the diad that do not change the
geometry reflect genuine d.o.f.

Before moving forward with this topic, let us explore the Hamiltonian
counterpart of the symmetry transformation~(\ref{trans}). Just as in GR,
the transformation~(\ref{trans}) can be regarded as generated by the
constraints. While the super-momentum and super-Hamiltonian constraints
generate the transformation of the space sector of the diad, the
transformation of the (pure gauge) timelike sector of the diad is generated
by the primary constraints. For this, let us compute
\begin{equation}
    \delta \mathbf{E}^{a}=\{\mathbf{E}^{a}(t,x),
    \int dx^{\prime }(\chi^{b}G_{b}^{(1)}+\xi ^{\nu }G_{\nu }^{(2)})\} \,,
    \label{transf}
\end{equation}
which yields
\begin{align}
    \delta E_{0}^{a}(t,x) &= \chi ^{a}(t,x) \,, \\
    \delta E_{1}^{a}(t,x) &= \int dy \delta (x-y)\int dx^{\prime }
    \left[ -\xi^{0}\frac{E}{4k} \eta^{ac}\pi _{c}^{1} \delta (x^{\prime }-y) -
    \xi ^{\nu}E_{\nu }^{a}~\partial _{x^{\prime }}\delta (x^{\prime }-y)\right] 
    \nonumber \\ &=
    -\xi ^{0}\frac{E}{4k}~\eta ^{ac}\pi _{c}^{1}+\partial _{x}(\xi ^{\nu}E_{\nu }^{a}) =
    \xi ^{0}T_{~01}^{a}+\partial _{x}(\xi ^{\nu }E_{\nu }^{a})
    \nonumber \\[1ex] &=
    \xi ^{\nu }\partial _{\nu }E_{1}^{a}+E_{\nu }^{a}~\partial _{1}\xi ^{\nu} =
    \left( \mathcal{L}_{\mathbf{\xi }}\mathbf{E}^{a}\right)_{1} \,.
\end{align}
Therefore, the space-time translation~(\ref{trans}) is obtained if the the 
parameters $\chi^a$ are fixed to be\footnote{The $E^a_0$-sector in Eq.~(\ref{transf}) would be avoided if it were removed from phase space. This can be achieved by promoting the $E^a_0$ variables to the role of (non-dynamical) Lagrange multipliers through a part integration in the Lagrangian action, as used in the ADM formalism of GR (cf.~Eq.~(\ref{accchil})).}
\begin{equation}
    \chi ^{a}(t,x)=\left( \mathcal{L}_{\mathbf{\xi }}\mathbf{E}^{a}\right) _{0} \,.
    \label{chi}
\end{equation}

To complete our understanding on how the space-time translations act in the 
phase space, let us now look at the transformation of the momenta:
\begin{equation}
    \mathbf{\pi}_{a} \mapsto \mathbf{\pi}_{a}+\delta \mathbf{\pi}_{a} = 
    \{\mathbf{\pi}_{a}(t,x),\int dx^{\prime }
    (\chi ^{b}G_{b}^{(1)}+\xi ^{\nu} G_{\nu }^{(2)})\} \,.
    \label{transfmomenta}
\end{equation}
Notice that we can ignore the first Poisson bracket even if $\chi^b$ were
replaced by~(\ref{chi}), because it would be weakly zero. Then
\begin{equation}
    \delta \pi^{0}_{a} \approx -\xi ^{0}(t,x) 
    \left(e^0_a E (-\frac{1}{8k} \eta^{dc}\pi_{d}^{1}\pi _{c}^{1}+k\Lambda) -\partial_1\pi^1_a\right)=-\xi ^{0} G^{(2)}_a\approx 0 \,,
\end{equation}
so that $\pi^0_a$ remains zero after the transformation. Moreover we have
\begin{align}
    \delta \pi^{1}_{a} & \approx 
    -\xi ^{0}(t,x) e^1_a  E (-\frac{1}{8k} \eta^{dc}\pi _{d}^{1}\pi _{c}^{1}+k\Lambda )+\xi ^{1 }(t,x) \partial_1\pi^1_a
    \nonumber \\ &= 
    -e^1_a \xi ^{\nu} G^{(2)}_\nu + \epsilon_{\mu\nu}e^\mu_a\xi^\nu E^b_0 \partial_1\pi^1_b\approx\epsilon_{\mu\nu}e^\mu_a\xi^\nu 
    E^b_0 \partial_1\pi^1_b \,.
    \label{transfpia1}
\end{align}
We are now ready to complete this discussion by considering the constraint algebra.

\subsection{The algebra of constraints}
\label{algebra} 

Now that the complete set of consistent constraints has been established in Section~\ref{consis}, we can continue and compute the algebra of
constraints. This is needed for the determination of the number of genuine d.o.f.
We will make repeated use of 
$\partial E/\partial E_{\mu }^{a}=Ee_{a}^{\mu }$ in the following
calculations. The commutators between primary constraints are zero,
\begin{equation}
    \{G_{a}^{(1)}(x),G_{b}^{(1)}(x^{\prime })\}=0 \,,
    \end{equation}%
while the commutators between the primary and secondary constraints are zero or
weakly zero. We have
\begin{align}
    \{G_{a}^{(1)}(x), G_{\nu }^{(2)}(x^{\prime })\} &=
    \Bigl\{\pi_{a}^{0}(x),\delta _{\nu }^{0} E(-\frac{1}{8k} 
    \eta ^{dc}\pi _{d}^{1}\pi_{c}^{1}+k\Lambda ) - E_{\nu }^{b} 
    \partial _{1}\pi _{b}^{1}\Bigr\}  
    \nonumber \\ &=
    \left[ \delta _{\nu }^{0}~\frac{\delta E}{\delta E_{0}^{a}}
    (\frac{1}{8k}\eta ^{dc}\pi _{d}^{1}\pi _{c}^{1}-k\Lambda ) + 
    \frac{\delta E_{\nu }^{b}}{\delta E_{0}^{a}}\partial _{1}\pi _{b}^{1}\right] \delta (x, x^{\prime })
    \nonumber \\ &=
    \delta _{\nu }^{0} \left[ e_{a}^{0} E(\frac{1}{8k} \eta^{dc}
    \pi_{d}^{1} \pi_{c}^{1}-k\Lambda) + 
    \partial _{1}\pi _{a}^{1}\right] ~\delta (x,x^{\prime })  
    \nonumber \\ &=
    -\delta _{\nu }^{0}~G_{a}^{(2)}~\delta (x, x^{\prime }) \,.
\end{align}
Thus the primary constraints $G_a^{(1)}$ are 1st class, since they commute
with all the constraints. First class constraints generate gauge
transformations. In our case $G_a^{(1)}=\pi^0_a$, so the gauge
transformations they generate affect the $E^a_0$. Thus the $E^a_0$ can
be chosen by means of a gauge fixing condition.

Let us now consider the algebra of secondary constraints. We might expect to
recognize the so called algebra of diffeomorphisms \cite{Dirac, DeWitt,Isham}. 
However, if this were the case, not only would the primary constraints be 1st class,
but the secondary constraints would be 1st class too. In that case no genuine d.o.f. would remain in this theory. In fact, four 1st class constraints would imply four gauge freedoms, since this theory has only four dynamical variables $E^a_\mu$ and all of them would be pure gauge variables. Therefore the algebra of super-momentum 
and super-Hamiltonian should evidence that they are instead 2nd class constraints.
This will now be shown.

The commutator of the super-momenta constraints is
\begin{align}
    \{G_{1}^{(2)}(x),~G_{1}^{(2)}(x^{\prime })\} &=
    \Bigl\{E_{1}^{c}~\partial _{1}\pi_{c}^{1},
    E_{1}^{d}~\partial _{1}\pi _{d}^{1}\Bigr\}  
    \nonumber \\[1ex] &=
    [\partial _{1}\pi _{a}^{1}]_{x} 
    [E_{1}^{a}]_{x^{\prime }} \partial_{x^{\prime }} \delta (x^{\prime },x)-[E_{1}^{a}]_{x}~\partial _{x}\delta(x,x^{\prime })
    [\partial _{1}\pi _{a}^{1}]_{x^{\prime }}
    \nonumber \\[1ex] &=
    G_{1}^{(2)}(x^{\prime })\partial _{x}\delta (x,x^{\prime}) -
    G_{1}^{(2)}(x)\partial_{x^{\prime }}\delta (x^{\prime },x) \,,
\end{align}
where $[\ ]_{x}$ stands for `evaluated at $x$' (see Appendix~\ref{Diracdelta} for more details). As can be seen, the super-momenta constitute a closed subalgebra, whose form is shared with other theories of gravity.

The commutator between super-Hamiltonian constraints is
\begin{multline}
    \{G_{0}^{(2)}(x),~G_{0}^{(2)}(x^{\prime })\} = 
    \Bigl\{E(\frac{1}{8k}\eta^{dc}\pi_{d}^{1}\pi_{c}^{1}-k\Lambda )+
    E_{0}^{a}\partial _{1}\pi_{a}^{1},E(\frac{1}{8k}\eta^{fe}
    \pi _{f}^{1}\pi _{e}^{1}-k\Lambda) +
    E_{0}^{b}~\partial_{1}\pi_{b}^{1}\Bigr\}  
    \\ =
    [e_{a}^{1}E(\frac{1}{8k}\eta ^{dc}\pi _{d}^{1}\pi_{c}^{1}-k\Lambda )]_{x} \left( [\frac{E}{4k}~\eta ^{ae}\pi_{e}^{1}]_{x^{\prime }}
    \delta (x^{\prime }, x)+[E_{0}^{a}]_{x^{\prime}}\partial_{x^{\prime }}
    \delta (x^{\prime }, x)\right)
    \\ 
    -\left( [\frac{E}{4k}~\eta ^{ac}\pi_{c}^{1}]_{x} \delta (x,x^{\prime }) + [E_{0}^{a}]_{x} \partial _{x}\delta (x, x^{\prime })\right)
    [e_{a}^{1}E(\frac{1}{8k} \eta^{fe}\pi_{f}^{1}\pi _{e}^{1}-k\Lambda)]_{x^{\prime }} = 0 \,.
\end{multline}
Here we used that $e_{a}^{1}E_{0}^{a} = \delta _{0}^{1}=0$, again, for more details see Appendix~\ref{Diracdelta}.

Finally, the commutator between the super-Hamiltonian and the 
super-momentum is
\begin{multline}
    \{G_{0}^{(2)}(x), G_{1}^{(2)}(x^{\prime })\} = 
    \Bigl\{E(\frac{1}{8k}\eta^{dc}\pi_{d}^{1}\pi_{c}^{1}-k\Lambda) + E_{0}^{a}\partial _{1}\pi_{a}^{1},
    E_{1}^{b}\partial_{1}\pi _{b}^{1}\Bigr\}  
    \\
    = [e_{b}^{1}E(\frac{1}{8k} \eta ^{dc}\pi _{d}^{1}\pi _{c}^{1} - 
    k\Lambda)]_{x} [E_{1}^{b}]_{x^{\prime }}~\partial _{x^{\prime }}\delta (x^{\prime },x) 
    -\frac{1}{4k} \eta ^{bc}[E~\pi _{c}^{1}]_{x}~[\partial _{1} \pi_{b}^{1}]_{x^{\prime }} \delta (x, x^{\prime }) 
    \\
    - 
    [E_{0}^{a}]_{x}~\partial _{x}\delta (x, x^{\prime })
    [\partial _{1}\pi_{a}^{1}]_{x^{\prime }}
    =
    G_{0}^{(2)}\partial_{x}\delta (x,x^{\prime })-\frac{E}{8k} 
    \partial_{1}(\eta^{ac}\pi_{a}^{1}\pi_{c}^{1})
    \delta (x, x^{\prime }) \,.
    \label{PB01}
\end{multline}
In this calculation the first term is characteristic of the algebra of diffeomorphisms. However, the second term prevents the closure of the algebra. Thus the super-momentum and the super-Hamiltonian do not weakly commute,  as would be expected in higher dimensions, and hence they are 2nd class constraints (cf.~the central charge in Ref.~\cite{Claudio}).  A pair of 2nd class constraints eliminates one d.o.f.; so a genuine d.o.f. still remains. In fact, we started with four dynamical variables $E^a_\mu$, two of which are eliminated by the two 1st class primary constraints $G^{(1)}_a$, and another one is eliminated by the pair of 2nd class, secondary constraints $G^{(2)}_\nu$. 

\subsection{Nature of the genuine degree of freedom}

The fact that neither the super-momentum nor the super-Hamiltonian constraints are 1st class means that space-time translations are not gauge transformations in two dimensions. This is true despite the theory being invariant under diffeomorphisms. Thus translations map solutions to \textit{physically different} solutions, that can be distinguished by the values the sole genuine d.o.f.~takes at each such solution.\footnote{In general relativity, instead, the result of the translation $\mathbf{g} \mapsto \mathbf{g} + \mathcal{L}_{\mathbf{\xi }}\mathbf{g}$ does not yield a physically new solution, rather it is the same geometry in different coordinates.}  Consequently, we should investigate the physical nature of this remaining degree of freedom. It cannot be a magnitude related exclusively to Riemannian geometry, since Jackiw-Teitelboim theory completely fixes the two-dimensional Riemannian geometry by fixing the curvature scalar.\footnote{The teleparallel framework is not based on a Riemannian geometry, even though both can be linked by means of the metric introduced in Eq.~(\ref{metric}). Teleparallel theories focus on the vielbein as a potential for the torsion field. Therefore, what we are really meaning in this sentence is that the genuine d.o.f.~is not (exclusively) related to the divergence of vector $T^\rho$ (see Eq.~(\ref{R2})).}  The d.o.f.~manifests itself through the integration constants of the booston field.

Let us examine the magnitude $Q = \eta^{ab}\pi^1_a\pi^1_b$ appearing in the central charge of Eq.~(\ref{PB01}). $Q$ does not belong to the pure gauge sector $(E^a_0,\pi^0_a)$ since it is constructed using variables of the dynamical sector $(E^a_1,\pi^1_a)$. Moreover, $Q$ is affected by space-time translations, as implied by Eq.~(\ref{transfpia1}). Therefore, $Q$ is a scalar quantity able to distinguish physically different solutions. Consequently it has to be related to the remaining d.o.f. 

According to Eq.~(\ref{pia1}) we have
\begin{equation}
    \label{elQ}
    Q=\eta^{ab}\pi^1_a\pi^1_b=16 k^2 E^{-2}\eta_{ab}T^a{}_{01}T^b{}_{01}= -8 k^2\,\mathbb{T}=-
    8 k^2\, T_\rho T^\rho \,,
\end{equation}
where the definition of $\mathbb{T}$ in Eq.~(\ref{action}) was used. It is becoming clear that the remaining d.o.f.~can be traced back to the Lagrangian density itself, see the action Eq.~(\ref{action}). This is not surprising after all, due to the fact that $\mathbb{T}$ is sensitive (not invariant) to local Lorentz transformations of the diad.

In the general solution we have shown in Section~\ref{solve} -- see Eqs.~(\ref{diad}), (\ref{diad1}), (\ref{dilatonUV}) and (\ref{booston}) -- the pure gauge variables $E^a_0$ have been gauge-fixed to be
\begin{equation}
    E^{\underline{0}}_0=E^{\underline{1}}_1\,, \qquad
    E^{\underline{1}}_0=E^{\underline{0}}_1\,.
    \label{gaugefix}
\end{equation}
On the other hand, the two dynamical variables $E^a_1$ have been written in terms of the two functions $\Phi$ and $\phi$
\begin{equation}
    E^{\underline{0}}_1=\exp[\Phi]\sinh\phi \,, \qquad
    E^{\underline{1}}_1=\exp[\Phi]\cosh\phi \,.
    \label{dynvar}
\end{equation}
In this form the volume $E$ depends exclusively on $\Phi$ since $E=\exp[2\Phi]$. The scalar $Q$ takes the form
\begin{align}
   Q &= 16 k^2 \exp[{-2\Phi}]\left[\left(\frac{\partial\phi}{\partial t}-\frac{\partial\Phi}{\partial x}\right)^2-\left(\frac{\partial\phi}{\partial x}-\frac{\partial\Phi}{\partial t}\right)^2\right]
   \nonumber \\ &=
   16 k^2 \exp[{-2\Phi}]~\frac{\partial(\phi-\Phi)}{\partial u}\frac{\partial(\phi+\Phi)}{\partial v}\,.
   \label{qencart}
\end{align}
Using the general solution (\ref{dilatonUV})--(\ref{booston}) we arrive at
\begin{equation}
    \label{elqgen}
    Q = -16 k^2 \frac{\left(u_0+\frac{\Lambda}{2}\alpha V\right)
    \left(v_0+\frac{\Lambda}{2}\beta U\right)}{(v_0 V-\beta)(u_0 U-\alpha)} \,.
\end{equation}
The sole genuine d.o.f. is expressed in $Q$ through the different values that the integration constants $\alpha, \beta$, $u_0, v_0$ can have in the solution (\ref{booston}) for the booston field. Remember they represent the origin $U=\alpha/u_o$, $V=\beta/v_o$ around which the booston $\phi$ performs its action, and the two straight lines of different slope where the booston is zero. 

We note, by examining Eq.~(\ref{PB01}), that a non-constant $Q$ is responsible for the non-closure of the algebra in order for the d.o.f.~to be active. Otherwise, the secondary constraints become 1st class, and no genuine d.o.f.~would be left. Thus, it is worth exploring the case when $Q$ is constant. This happens under the condition
\begin{equation}
    2 u_0 v_0 + \alpha\beta\Lambda = 0
    \quad\Rightarrow\quad
    Q = 8k^2 \Lambda\,.
   \label{onlyc}
\end{equation}
Then the momenta (\ref{pia1}) for the solution (\ref{dilatonUV})--(\ref{booston}) also become constant. In fact, let us satisfy the condition (\ref{onlyc}) by choosing 
\begin{equation}
    u_0=-\text{sign}(\Lambda)\ e^\gamma\beta\sqrt{2\vert\Lambda\vert}\,, \qquad
    v_0= e^{-\gamma}\alpha\sqrt{2\vert\Lambda\vert}\,,
    \label{replace}
\end{equation}
where $\gamma$ is an arbitrary parameter. Thus the momenta $\pi^1_a$ become
\begin{equation}
    \pi^1_{\underline 0} = \text{sign}(\Lambda)\ 2k\,\sqrt{2\vert\Lambda\vert} \cosh\gamma\,, \qquad
    \pi^1_{\underline 1} = \text{sign}(\Lambda)\ 2k\,\sqrt{2\vert\Lambda\vert} \sinh\gamma\,.
\end{equation}
Since the momenta $\pi^1_a$ are constant they are not affected by space-time translations, as can be checked in Eq.~(\ref{transfpia1}), where one finds $\delta \pi^1_a \approx 0$. More importantly, the second term in the super-Hamiltonian,
\begin{equation}
    \partial_1(E^a_0\pi^1_a)=k\, \frac{2 u_0 v_0+\alpha\beta\Lambda}{(u_0 U-\alpha)(v_0 V-\beta)(1+\frac{\Lambda}{2}UV)}\,,
    \label{diff}
\end{equation}
becomes zero if Eq.~(\ref{onlyc}) is satisfied. Therefore the super-Hamiltonian is equal to the canonical Hamiltonian density, which is typical of theories where the space-time translations return to be part of the diffeomorphisms. This conclusion seems to be confirmed by the form the booston~(\ref{booston}) acquires after using the substitution (\ref{replace}). This gives 
\begin{equation}
    \phi=\log\left\vert\frac{1-\sqrt{\vert\Lambda\vert/2} \,e^{-\gamma}\, V\alpha/\beta}{1+\text{sign}(\Lambda) \sqrt{\vert\Lambda\vert/2} \,e^{\gamma}\, U\beta/\alpha}\right\vert\,.
\end{equation}
Despite the apparent freedom associated with the choice of the integration constants $\gamma$, $\alpha/\beta$, the booston is completely determined, because these constants can be absorbed by the only null-coordinate transformations we are allowed to make (see Note \ref{foot} for details) 
\begin{equation}
    e^{\gamma} U\beta/\alpha \rightarrow U\,,\qquad
    e^{-\gamma} V\alpha/\beta \rightarrow V\,.
\end{equation}
Therefore no genuine d.o.f.~is left when $Q$ is constant, since no free integration constants remain in the booston, which is the result we wished to establish.

Note that $Q$ will be null if and only if $\Lambda=0$, in which case at least one of $u_0$ or $v_0$ must also be zero in Eq.~(\ref{onlyc}). This is fulfilled not only by the global booston having $u_0=v_o=0$, but also by the \textit{in} and \textit{out} modes 
\begin{alignat}{3}
    \mathbf{E}^{{\underline{0}}}_{in} &= 
    \frac{\mathbf{d}U+\mathbf{d}V}{2(v_{0}\,V+\beta)} \,,&\qquad 
    \mathbf{E}^{{\underline{1}}}_{in} &= 
    \frac{\mathbf{d}U-\mathbf{d}V}{2(v_{0}\,V+\beta)} \,,&\qquad
    &(u_{0}=0,\Lambda=0) \,,
    \label{modoout} \\[1ex]
    \mathbf{E}^{{\underline{0}}}_{out} &= 
    \frac{\mathbf{d}U+\mathbf{d}V}{2( u_{0}\,U+\alpha)} \,,&\qquad \mathbf{E}^{{\underline{1}}}_{out} &= 
    \frac{\mathbf{d}U-\mathbf{d}V}{2(\,u_{0}\,U+\alpha)} \,,&\qquad
    &(v_{0}=0,\Lambda=0) \,.
    \label{modoin}
\end{alignat}
According to Eq.~(\ref{elQ}), the value of $Q$ as coming from Eq.~(\ref{onlyc}), implies $\mathbb{T}=0$ (if $\Lambda=0$), or $\mathbb{T}=-\Lambda$ otherwise. In the first case, the on-shell action is identically null, whereas in the second it becomes
\begin{equation}
    S[\mathbf{E}^{a}] = k\int E\, 
    (\,\mathbb{T} - \Lambda ) \, dx^{0} dx^{1} = -2k \int E 
    \Lambda \, dx^{0} dx^{1}\,. 
    \label{actiontopcom}
\end{equation}
Because of the JT field equation $R=\Lambda$, the right-hand side of the previous action can be understood in terms of the Euler characteristic of the manifold which is defined by 
\begin{equation}
    \chi_{E} := \frac{1}{4\pi} \int\limits_{\mathscr{M}} R \sqrt{|g|}\, d^2 x \,, 
\end{equation}
so that our action takes the topological form
\begin{equation}
    S[\mathbf{E}^{a}] = -8\pi k\, \chi_{E}\, .  
    \label{actiontop}
\end{equation}
Therefore, the absence of the degree of freedom, captured in the condition $Q=constant$, is consistent with the fact that the action becomes purely topological, regardless of the different values of the integration constants that can be combined to satisfy the condition~(\ref{onlyc}). This matches the standard discussions in GR where the two dimensional action is topological. 

\section{Conclusions}

General Relativity is an intrinsically four dimensional theory formulated using the language of differential geometry. It gives ten coupled non-linear equations which contain two propagating degrees of freedom which are associated with the two polarisations of gravitational waves. Direct attempts at quantising the theory have not yielded success, which motivated the study of models in lower dimensions where the equations are considerably simpler. Neither three dimensional nor two dimensional gravity contain propagating degrees of freedom. 2D GR is well-known to be topological and some extra structure needs to be introduced to make the theory non-trivial.

We present a new approach to this problem by starting out from the Teleparallel Equivalent of GR, known as TEGR. In three and four spacetime dimensions this theory is completely equivalent to GR but is formulated using tetrads instead of metric as the fundamental field, and employs the Weitzenböck connection instead of the standard Levi-Civita one. When similar ideas are considered in two dimensions, one arrives naturally at a non-trivial two dimensional theory based on the torsion produced by the Weitzenböck connection, which constitutes the first result reviewed in this work. That this is indeed possible is a quirk of the torsion tensor and its irreducible composition. In two dimensions, only a vector torsion part is allowed and the norm of this vector can naturally be considered as the action of a gravitational toy model, which, in general, has no direct link to a topological quantity. 

We demonstrate, using the Hamiltonian constraint analysis, that the resulting theory has one genuine degree of freedom, in general. We identify the corresponding quantity $Q=-8 k^2\,\mathbb{T}$ and discuss some of its properties. 

It is particularly interesting to note that the starting action is invariant under arbitrary coordinate transformations and global Lorentz transformations, while the Hamiltonian analysis shows that the spacetime translations are symmetries of the theory, but they are not gauge transformations (they are not 1st class). Normally they are incorporated into the general coordinate transformations, but the theory here discussed seems to link the breaking of the local Lorentz invariance with the appearance of translations mapping one solution into a physically different solution, i.e., translations are not part of diffeomorphisms in two spacetime dimensions. This finding, revealed through the Hamiltonian analysis, constitutes the core of our article.  

Our approach yields a theory that has close links with Jackiw-Teitelboim gravity. In particular, our theory is naturally derived using a variational approach, something that is lacking in the standard JT formulation. Since our action is quadratic in the torsion vector, and recalling that the torsion vector contains first derivatives of the tetrad fields, we are dealing with a Yang-Mills type theory. In such theories the action always contains squares of field strengths, and field strengths are the first derivatives of the potentials. JT gravity appears, then, as a local Lorentz invariant sector of the torsion-based theory here proposed. 

Given that our toy model contains a single d.o.f., it is rather natural to consider generalised models based on the simplest one discussed here. The key quantity in the action is the torsional scalar $\mathbb{T}$. Consequently, as briefly mentioned in \cite{Andronikos}, one can consider
\begin{equation}  
    \label{accionefe}
    S_{f} \propto \int f(\mathbb{T}) E\, d^2x\,,
\end{equation}
which contains our model when choosing $f(\mathbb{T})=\mathbb{T}-\Lambda$. One can now speculate that this model will contain at least one additional degree of freedom due to the presence of the function $f$. However, this is merely a lower bound for the following reason. Just as in GR, in TEGR one finds two d.o.f., however, $f(T)$ gravity in four dimensions\footnote{We recall that $T$ is the Weitzenböck scalar introduced in Eq.~(\ref{R1}). For a review of many of the vast number of results in the area, including some discrepancies in the counting of d.o.f., we refer the reader to Ref.~\cite{Review}.} has up to five d.o.f., which is considerably more than one might expect by simply including a new function. This has to do with the fact that $f(\mathbb{T})$, just as the particular choice $f(\mathbb{T})=\mathbb{T}-\Lambda$ here considered, also breaks local Lorentz invariance and it becomes a non-trivial task to establish the number of propagating degrees of freedom according to the Hamiltonian analysis. One can also relate this to the hardly understood \emph{remnant symmetries}, which are local Lorentz transformations leaving $T$ invariant up to boundary terms \cite{grupo}. Thus two dimensional $f(\mathbb{T})$ gravity is an excellent toy model to be considered further.  

Since the quantity $\mathbb{T}=T^\rho T_\rho$ is constructed from the torsion vector, one can also consider other gravitational toy models using, for example, the symmetric matrix $\mathbf{F}=F_{\alpha\beta} := T_\alpha T_\beta$. This is motivated by torsional Born-Infeld determinantal gravity models, which are of the form
\begin{equation}  
    \label{ac2d2}
    S_{\rm BI} \propto \int \Big[
    \sqrt{\det(\mathbf{I} + 2\lambda^{-1} \mathbf{F})}-\Lambda
    \Big] E\, d^2x \,,
\end{equation}
see Refs.~\cite{bidet} and~\cite{Vatu}. Here $\mathbf{I}$ stands for the identity matrix and $\lambda$ is the Born-Infeld parameter. Of course, one could consider arbitrary functions depending on the determinant as an argument. As mentioned before, when considering such models it is not clear how many d.o.f.~such a theory will have. We can speculate that the lower bound is again one, and remind the reader that the upper bound is four, the number of independent component of the diad, which is the dynamical variable of the theory.

It is rather remarkable that two dimensional gravitational models based on the teleparallel framework offer such a rich dynamical structure not seen in its GR counterpart. This route to study such models has been largely overlooked and it is hoped that this contribution will help to initiate some progress in the field.   

\subsection*{Acknowledgements.} 

FF thanks the Department of Mathematics at UCL where part of this work was done. He also  acknowledges financial support by the UCL MAPS Visiting Fellowship Scheme. RF and FF are members of \emph{Carrera del Investigador Cient\'{i}fico} (CONICET). This work has been partially supported by CONICET, Instituto Balseiro (UNCUYO) and Universidad de Buenos Aires.

\appendix

\section{Additional details for the Hamiltonian analysis}

\subsection{Consistent evolution of \texorpdfstring{$G^{(2)}_a$}{G(2)a}}
\label{Ga}

To prove the consistency of the evolution of the secondary constraints $G^{(2)}_a$, we have to compute the Poisson bracket 
\begin{equation}
    \{e_{a}^{0}\left(
    H_{C}-\pi _{b}^{1}~\partial _{1}E_{0}^{b}\right) -\partial _{1}\pi
    _{a}^{1}~,\int dx^{\prime }H_{C}\}\label{PBGa}\, ,
\end{equation}
which enters in Eq.~(\ref{evolveGa}). We will use $e_{b}^{\lambda}E_{\lambda }^{c}=\delta _{b}^{c}$ to obtain $\delta e_{a}^{0}/\delta
E_{\mu }^{c}$, this is
\begin{equation}
    \delta e_{b}^{\lambda }~E_{\lambda }^{c}=-e_{b}^{\lambda }~\delta E_{\lambda
    }^{c}
    \qquad\Rightarrow\qquad
    \frac{\delta e_{b}^{\nu }}{\delta E_{\mu}^{c}}=-e_{b}^{\mu }e_{c}^{\nu }\,.
\end{equation}%
Then, by using Eq.~(\ref{varHp}) we get
\begin{eqnarray}
    0 &\approx &\{e_{a}^{0}\left( H_{C}-\pi _{b}^{1}~\partial
    _{1}E_{0}^{b}\right) -\partial _{1}\pi _{a}^{1}~,\int dx^{\prime }H_{C}\}
    \notag \\
    &=&-\int dy~\delta (x-y)~e_{a}^{1}e_{c}^{0}~(H_{C}-\pi _{b}^{1}~\partial
    _{1}E_{0}^{b})~\left[ -\frac{E}{4k}~\eta ^{dc}\pi _{d}^{1}+\partial
    _{1}E_{0}^{c}\right] _{y}  \notag \\
    &&+\int dy~\delta (x-y)~e_{a}^{0}~\partial _{1}E_{0}^{b}~\left[ e_{b}^{1}E~(-%
    \frac{1}{8k}~\eta ^{dc}\pi _{d}^{1}\pi _{c}^{1}+k~\Lambda )\right] _{y}
    \notag \\
    &&+\int dy~\partial _{x}\delta (x-y)~\left[ e_{a}^{1}E~(-\frac{1}{8k}~\eta
    ^{dc}\pi _{d}^{1}\pi _{c}^{1}+k~\Lambda )\right] _{y}  \notag \\
    &=&-e_{a}^{1}e_{c}^{0}~E~(-\frac{1}{8k}~\eta ^{eb}~\pi _{e}^{1}\pi
    _{b}^{1}+k~\Lambda )~\left( -\frac{E}{4k}~\eta ^{dc}\pi _{d}^{1}+\partial
    _{1}E_{0}^{c}\right)  \notag \\
    &&+e_{a}^{0}~e_{b}^{1}~\partial _{1}E_{0}^{b}~E~(-\frac{1}{8k}~\eta ^{dc}\pi
    _{d}^{1}\pi _{c}^{1}+k~\Lambda )+\partial _{1}\left[ e_{a}^{1}~E~(-\frac{1}{%
    8k}~\eta ^{dc}\pi _{d}^{1}\pi _{c}^{1}+k~\Lambda )\right]\, .  \label{result1}
    \end{eqnarray}%
This result should be examined on the constraint surface where, according to Eq.~(\ref{super}), it is
\begin{equation}
    E~(-\frac{1}{8k}~\eta ^{dc}\pi _{d}^{1}\pi _{c}^{1}+k~\Lambda
    )=E_{0}^{b}~\partial _{1}\pi _{b}^{1} \,,
    \qquad 
    0=E_{1}^{b}~\partial _{1}\pi_{b}^{1}\, .
\end{equation}
In particular, we have%
\begin{equation*}
    \partial _{1}\pi _{a}^{1}=\delta _{a}^{b}~\partial _{1}\pi
    _{b}^{1}=(e_{a}^{0}~E_{0}^{b}+e_{a}^{1}~E_{1}^{b})~\partial _{1}\pi
    _{b}^{1}\approx e_{a}^{0}~E_{0}^{b}~\partial _{1}\pi _{b}^{1}\, .
\end{equation*}%
Thus the first term in Eq.~(\ref{result1}) is%
\begin{eqnarray}
    -e_{a}^{1}e_{c}^{0}~E_{0}^{b}~\partial _{1}\pi _{b}^{1}~\left( -\frac{E}{4k}%
    ~\eta ^{dc}\pi _{d}^{1}+\partial _{1}E_{0}^{c}\right) &\approx
    &-e_{a}^{1}~\partial _{1}\pi _{c}^{1}~\left( -\frac{E}{4k}~\eta ^{dc}\pi
    _{d}^{1}+\partial _{1}E_{0}^{c}\right)  \notag \\
    &=&\frac{E}{8k}~e_{a}^{1}~\partial _{1}\left( \eta ^{dc}\pi _{c}^{1}\pi
    _{d}^{1}\right) -e_{a}^{1}~\partial _{1}\pi _{c}^{1}~\partial _{1}E_{0}^{c}\, .
\end{eqnarray}%
The second term in Eq.~(\ref{result1}) is%
\begin{equation}
    e_{a}^{0}~e_{b}^{1}~\partial _{1}E_{0}^{b}~E~(-\frac{1}{8k}~\eta ^{dc}\pi
    _{d}^{1}\pi _{c}^{1}+k~\Lambda )\approx e_{a}^{0}~e_{b}^{1}~\partial
    _{1}E_{0}^{b}~E_{0}^{c}~\partial _{1}\pi _{c}^{1}\approx e_{b}^{1}~\partial
    _{1}E_{0}^{b}~\partial _{1}\pi _{a}^{1}=-E_{0}^{b}~\partial
    _{1}e_{b}^{1}~\partial _{1}\pi _{a}^{1}\, .
\end{equation}%
The third term in Eq.~(\ref{result1}) is%
\begin{eqnarray}
    \partial_{1}\left[ e_{a}^{1}E(-\frac{1}{8k}~\eta ^{dc}\pi _{d}^{1}\pi
    _{c}^{1}+k~\Lambda )\right] &=&-\frac{E}{8k}~e_{a}^{1}~\partial _{1}\left(
    \eta ^{dc}\pi _{c}^{1}\pi _{d}^{1}\right) +(-\frac{1}{8k}~\eta ^{dc}\pi
    _{d}^{1}\pi _{c}^{1}+k~\Lambda )~\partial _{1}(e_{a}^{1}~E)  \notag \\
    &\approx &-\frac{E}{8k}~e_{a}^{1}~\partial _{1}\left( \eta ^{dc}\pi
    _{c}^{1}\pi _{d}^{1}\right) +E_{0}^{b}~\partial _{1}\pi
    _{b}^{1}~E^{-1}\partial _{1}(e_{a}^{1}~E)  \notag \\
    &=&-\frac{E}{8k}~e_{a}^{1}~\partial _{1}\left( \eta ^{dc}\pi _{c}^{1}\pi
    _{d}^{1}\right) +E_{0}^{b}~\partial _{1}\pi _{b}^{1}~(\partial
    _{1}e_{a}^{1}+e_{a}^{1}~e_{c}^{\nu }~\partial _{1}E_{\nu }^{c})\, .
\end{eqnarray}%
Therefore, Eq.~(\ref{result1}) is
\begin{equation}
    0\approx -e_{a}^{1}~\partial _{1}\pi _{c}^{1}~\partial
    _{1}E_{0}^{c}-E_{0}^{b}~\partial _{1}e_{b}^{1}~\partial _{1}\pi
    _{a}^{1}+E_{0}^{b}~\partial _{1}\pi _{b}^{1}~(\partial
    _{1}e_{a}^{1}+e_{a}^{1}~e_{c}^{\nu }~\partial _{1}E_{\nu }^{c})\, .
\end{equation}%
To verify that these equations do not lead to new constraints but they are
satisfied on the constraint surface, let us contract them with $E_{\lambda
}^{a}$:
\begin{equation}
0\approx -\delta _{\lambda }^{1}~\partial _{1}\pi _{c}^{1}~\partial
_{1}E_{0}^{c}-E_{0}^{b}~\partial _{1}e_{b}^{1}~E_{\lambda }^{a}~\partial
_{1}\pi _{a}^{1}+E_{0}^{b}~\partial _{1}\pi _{b}^{1}~(E_{\lambda
}^{a}~\partial _{1}e_{a}^{1}+\delta _{\lambda }^{1}~e_{c}^{\nu }~\partial
_{1}E_{\nu }^{c})\, .
\end{equation}%
For $\lambda =0$ we have an identity. For $\lambda =1$, one obtains
\begin{eqnarray}
0 &\approx &-\partial _{1}\pi _{c}^{1}~\partial
_{1}E_{0}^{c}-E_{0}^{b}~\partial _{1}e_{b}^{1}~E_{1}^{a}~\partial _{1}\pi
_{a}^{1}+E_{0}^{b}~\partial _{1}\pi _{b}^{1}~(E_{1}^{a}~\partial
_{1}e_{a}^{1}+e_{c}^{\nu }~\partial _{1}E_{\nu }^{c})  \notag \\
&\approx &-\partial _{1}\pi _{c}^{1}~\partial
_{1}E_{0}^{c}+E_{0}^{b}~\partial _{1}\pi _{b}^{1}~e_{c}^{0}~\partial
_{1}E_{0}^{c}\approx -\partial _{1}\pi _{c}^{1}~\partial
_{1}E_{0}^{c}+\partial _{1}\pi _{c}^{1}~\partial _{1}E_{0}^{c}=0~.
\end{eqnarray}
Therefore, the Poisson bracket (\ref{PBGa}) is weakly zero.

\subsection{Commutators}
\label{Diracdelta} 

The commutators $\{G_{\nu
}^{(2)}(x),~G_{\nu }^{(2)}(x^{\prime })\}$ in \S \ref{algebra} are
distributions antisymmetric in $(x,x^{\prime })$. They contain combinations
of terms with the generic form
\begin{equation}
\lbrack g_{a}]_{x}~[h^{a}]_{x^{\prime }}~\partial _{x}\delta (x,x^{\prime
})-[g_{a}]_{x^{\prime }}~[h^{a}]_{x}~\partial _{x^{\prime }}\delta
(x^{\prime },x)~.  \label{combination}
\end{equation}%
To understand how this distribution works, let us apply it to a function $%
f(x^{\prime })$:
\begin{eqnarray}
&&\int dx^{\prime }~f(x^{\prime })~\left( [g_{a}]_{x}[h^{a}]_{x^{\prime
}}~\partial _{x}\delta (x,x^{\prime })-[g_{a}]_{x^{\prime
}}[h^{a}]_{x}~\partial _{x^{\prime }}\delta (x^{\prime },x)\right)   \notag
\\
&=&[g_{a}]_{x}~\partial _{x}\int dx^{\prime }~[f~h^{a}]_{x^{\prime }}~\delta
(x,x^{\prime })+[h^{a}]_{x}\int dx^{\prime }~\partial _{x^{\prime
}}[f~g_{a}]_{x^{\prime }}~\delta (x^{\prime },x)  \notag \\
&&  \notag \\
&=&[g_{a}~\partial _{x}(f~h^{a})+h^{a}~\partial
_{x}(f~g_{a})]_{x}=[2g_{a}h^{a}~\partial _{x}f~+f~\partial
_{x}(g_{a}h^{a})]_{x}\, .
\end{eqnarray}%
Therefore, it is licit to rewrite the distribution (\ref{combination}) as%
\begin{eqnarray*}
\lbrack g_{a}]_{x}~[h^{a}]_{x^{\prime }}~\partial _{x}\delta (x,x^{\prime
})-[g_{a}]_{x^{\prime }}~[h^{a}]_{x}~\partial _{x^{\prime }}\delta
(x^{\prime },x) &=&[g_{a}h^{a}]_{x^{\prime }}~\partial _{x}\delta
(x,x^{\prime })-[g_{a}h^{a}]_{x}~\partial _{x^{\prime }}\delta (x^{\prime
},x)~, \\
&=&[g_{a}h^{a}]_{x}~\partial _{x}\delta (x,x^{\prime
})-[g_{a}h^{a}]_{x^{\prime }}~\partial _{x^{\prime }}\delta (x^{\prime },x)\, ,
\end{eqnarray*}%
as can be easily verified by repeating the procedure to obtain the same
result.

In the case of $\{G_{0}^{(2)}(x),~G_{0}^{(2)}(x^{\prime })\}$, $%
g_{a}h^{a}$ is zero because it involves the factor $e_{a}^{1}E_{0}^{a}=\delta
_{0}^{1}=0$. Besides, $\{G_{0}^{(2)}(x),~G_{0}^{(2)}(x^{\prime })\}$ also
contains a combination of the form%
\begin{equation}
\lbrack g_{a}]_{x}~[h^{a}]_{x^{\prime }}~\delta (x,x^{\prime
})-[g_{a}]_{x^{\prime }}~[h^{a}]_{x}~\delta (x^{\prime },x)~,
\end{equation}%
that results in zero when applied to an arbitrary function:
\begin{eqnarray}
&&\int dx^{\prime }~f(x^{\prime })~\left( [g_{a}]_{x}[h^{a}]_{x^{\prime
}}~\delta (x,x^{\prime })-[g_{a}]_{x^{\prime }}[h^{a}]_{x}~\delta (x^{\prime
},x)\right)   \notag \\
&=&[g_{a}]_{x}~\int dx^{\prime }~[f~h^{a}]_{x^{\prime }}~\delta (x,x^{\prime
})-[h^{a}]_{x}\int dx^{\prime }~[f~g_{a}]_{x^{\prime }}~\delta (x^{\prime
},x)=0~.
\end{eqnarray}

\end{document}